\documentclass[11pt]{article}
\usepackage{amssymb}
\usepackage{amssymb}
\usepackage{amsmath}

\usepackage{epsfig}
\usepackage{graphicx}

\numberwithin{equation}{section}

\def\be{\begin{equation}}
\def\ee{\end{equation}}
\def\ba{\begin{align}}
\def\ea{\end{align}}

\def\p{\partial}

\def\yboxit#1#2{\vbox{\hrule height #1 \hbox{\vrule width #1
\vbox{#2}\vrule width #1 }\hrule height #1 }}
\def\fillbox#1{\hbox to #1{\vbox to #1{\vfil}\hfil}}
\def\ybox{{\lower 1.3pt \yboxit{0.4pt}{\fillbox{8pt}}\hskip-0.2pt}}
\def\comments#1{}

\def\jb{{\bar j}}

\def\mb{{\bar m}}
\def\nb{{\bar n}}

\def\cb{{\bar c}}

\def\zb{{\bar z}}

\def\p{\partial}

\def\half{\frac{1}{ 2}}

\def\diag{{\rm diag}}

\def\ket#1{|#1\rangle}

\def\vev#1{\langle{#1}\rangle}

\def\cL{{\cal L}}

\def\cN{{\cal N}}
\def\cO{{\cal O}}

\def\cV{{\cal V}}

\def\CN{{\cal N}}

\def\CV{{\cal V}}
\def\CO{{\cal O}}

\def\CB{{\cal B}}

\def\CV{{\cal V }}

\def\II{\relax{I\kern-.10em I}}

\def\IZ{\relax\ifmmode\mathchoice
{\hbox{\cmss Z\kern-.4em Z}}{\hbox{\cmss Z\kern-.4em Z}}
{\lower.9pt\hbox{\cmsss Z\kern-.4em Z}}
{\lower1.2pt\hbox{\cmsss Z\kern-.4em Z}}\else{\cmss Z\kern-.4em
Z}\fi}
\def\IB{\relax{\rm I\kern-.18em B}}
\def\IC{{\relax\hbox{$\inbar\kern-.3em{\rm C}$}}}
\def\ID{\relax{\rm I\kern-.18em D}}
\def\IE{\relax{\rm I\kern-.18em E}}
\def\IF{\relax{\rm I\kern-.18em F}}
\def\IG{\relax\hbox{$\inbar\kern-.3em{\rm G}$}}
\def\IGa{\relax\hbox{${\rm I}\kern-.18em\Gamma$}}
\def\IH{\relax{\rm I\kern-.18em H}}
\def\II{\relax{\rm I\kern-.18em I}}
\def\IK{\relax{\rm I\kern-.18em K}}
\def\IN{\relax{\rm I\kern-.18em N}}
\def\IP{\relax{\rm I\kern-.18em P}}
\def\inbar{\,\vrule height1.5ex width.4pt depth0pt}

\def\p{\partial}

\font\cmss=cmss10 \font\cmsss=cmss10 at 7pt
\def\IR{\relax{\rm I\kern-.18em R}}

\def\CN{{\cal N}}

\def\lp10{l_P^{10}}
\def\lp11{l_P^{11}}
\def\R11{R_{11}}

\def\zb {\bar{z}}
\def\xb {\bar{x}}
\def\mb {\bar{m}}
\def\Jb {\bar{J}}
\def\Xb {\bar{X}}
\def\Hb {\bar{H}}
\def\Gb {\bar{G}}
\def\jb {\bar{j}}
\def\psib{\bar{\psi}}

\def\G{\Gamma}

\font\manual=manfnt
\def\dbend{\lower3.5pt\hbox{\manual\char127}}

\def\IZ{\relax\ifmmode\mathchoice
{\hbox{\cmss Z\kern-.4em Z}}{\hbox{\cmss Z\kern-.4em Z}}
{\lower.9pt\hbox{\cmsss Z\kern-.4em Z}} {\lower1.2pt\hbox{\cmsss
Z\kern-.4em Z}}\else{\cmss Z\kern-.4em Z}\fi}
\def\half {\frac{1}{ 2}}

\def\p{\partial}

\def\bar{\overline}

\def\CN{{\cal N}}

\def\rt2{\sqrt{2}}
\def\irt2{\frac{1}{\sqrt{2}}}

\def\t{\tilde}

\def\s{\sigma}
\def\b{\beta}

\font\cmss=cmss10
\font\cmsss=cmss10 at 7pt
\def\IL{\relax{\rm I\kern-.18em L}}
\def\IH{\relax{\rm I\kern-.18em H}}
\def\IR{\relax{\rm I\kern-.18em R}}
\def\inbar{\vrule height1.5ex width.4pt depth0pt}
\def\IC{\relax\hbox{$\inbar\kern-.3em{\rm C}$}}
\def\rlx{\relax\leavevmode}
\def\ZZ{\rlx\leavevmode\ifmmode\mathchoice{\hbox{\cmss Z\kern-.4em Z}}
 {\hbox{\cmss Z\kern-.4em Z}}{\lower.9pt\hbox{\cmsss Z\kern-.36em Z}}
 {\lower1.2pt\hbox{\cmsss Z\kern-.36em Z}}\else{\cmss Z\kern-.4em
 Z}\fi}
\def\IZ{\relax\ifmmode\mathchoice
{\hbox{\cmss Z\kern-.4em Z}}{\hbox{\cmss Z\kern-.4em Z}}
{\lower.9pt\hbox{\cmsss Z\kern-.4em Z}}
{\lower1.2pt\hbox{\cmsss Z\kern-.4em Z}}\else{\cmss Z\kern-.4em
Z}\fi}

\def\zb {\bar{z}}
\def\nb {\bar{n}}
\def\cb {\bar{c}}

\def\G{\Gamma}

\font\manual=manfnt
\def\dbend{\lower3.5pt\hbox{\manual\char127}}

\def\IZ{\relax\ifmmode\mathchoice
{\hbox{\cmss Z\kern-.4em Z}}{\hbox{\cmss Z\kern-.4em Z}}
{\lower.9pt\hbox{\cmsss Z\kern-.4em Z}} {\lower1.2pt\hbox{\cmsss
Z\kern-.4em Z}}\else{\cmss Z\kern-.4em Z}\fi}
\def\half {\frac{1}{ 2}}

\def\bar{\overline}

\def\rt2{\sqrt{2}}
\def\irt2{\frac{1}{\sqrt{2}}}

\def\t{\tilde}

\def\s{\sigma}

\begin{document}

\vskip-20pt

\begin{flushright}
IC/2005/120\\
LPTENS-05/37\\
\end{flushright}

\vskip 10pt

\begin{center}

{\huge {Topological Cigar and the $c=1$ String : \\ 
\vskip4pt
Open and Closed}
}
\end{center}

\begin{center}

{\large {Sujay K. Ashok$^{a}$, \
  Sameer Murthy$^{b}$
  and Jan Troost$^{c}$
}}

\end{center}

\date{}

\begin{center}
$^{a}$Perimeter Institute for Theoretical Physics\\
Waterloo, Ontario, ON N$2$L$2$Y$5$, Canada \\
\vskip2pt
$^{b}$Abdus Salam International Center for Theoretical Physics \\
Strada Costiera 11, Trieste, $34014$, Italy \\
\vskip2pt
$^{c}$Laboratoire de Physique Th\'eorique, Ecole Normale Sup\'erieure \\
Unit\'e mixte du CNRS et de l'Ecole
Normale Sup\'erieure
\\
$24$, Rue Lhomond Paris $75005$, France
\end{center}

\begin{abstract}
We clarify some aspects of the 
map between the $c=1$ string theory at self-dual radius and the topologically twisted cigar at level one. 
We map the ZZ and FZZT D-branes in the $c=1$ string theory at self dual radius to the localized and extended branes in the topological theory on the cigar. 
We show that the open string spectrum on the branes in the two theories are in correspondence with each other,  and their two point correlators are equal.
We also find a representation of an extended $\CN=2$ algebra on the worldsheet which incorporates higher spin currents in terms of asymptotic variables on the cigar.
\end{abstract}

%\date{}

\section{Introduction}

It was shown in \cite{Mukhi} that the topological string theory on the coset $SL(2,\IR)_{1}/U(1)$ (the cigar at supersymmetric level one) is equivalent to the $c=1$ non-critical bosonic string at radius $R=\sqrt{\alpha'}$.\footnote{Recent work \cite{Takayanagi} maps the topological cigar at arbitrary level $k$ to bosonic non-minimal $c<1$ matter coupled to Liouville.} In the last few years, there has been a lot of progress \cite{Teschnerpapers, Ponsot, Ribault, Stoyanovsky, RibaultTeschner} towards the full solution of the $SL(2,\IR)/U(1)$ coset theory. Using these results, a large class of closed string correlation functions of the two theories were explicitly checked to agree \cite{Nakamura}. We extend this map and identify the respective D-branes of the two theories with each other, by observing that the one-point functions defining boundary states in the $c=1$ string coincide with the topological couplings of boundary states in the cigar, and that the spectrum and two point functions of the open strings on these branes are equal.

Before discussing the open string theories, we clarify aspects of the
equivalence of the cohomologies of the topological coset and the $c=1$ bosonic
string theory. Using the analysis of null vectors in Verma modules of affine
$SL(2,\IR)$, we identify the origin of the bosonic string cohomology in the
supercoset Hilbert space\footnote{The observation that an $SL(2,\IR)$ current
  algebra appears in the context of light-cone $2$D gravity was already made in \cite{Polyakov,Knizhnik}. There, the relation between $c=1$ observables and degenerate representations of $SL(2,\IR)$ is also mentioned briefly.}. We thereby reproduce and clarify the relation between the analysis of the cohomology as described in the appendix of \cite{Mukhi} (by E. Frenkel) and the identification of certain explicit representatives of the $c=1$ string cohomology in \cite{Mukhi}. 

It was observed in \cite{Mukhi} that {\it all} values of the $SL(2,\IR)$ quantum number $j$ such that $2j \in \IZ$ have to be included for the map of closed string operators between the $c=1$ string and the coset to work. 
However it was shown in \cite{Maldacena} that the Hilbert space for string theory on the parent $SL(2,\IR)$ theory  is obtained by restricting $j$ to be in the range\footnote{In our conventions, where $k$ here is the supersymmetric level.} $-\half \ge j > -\frac{k+1}{2}$  while allowing for all spectrally flowed sectors. 
The spectrum of the coset theory is obtained by descent from the same (restricted) parent Hilbert space\cite{Hanany}. For $k=1$, this Hilbert space for the discrete representations is given by $\sum_w D^{+,w}_{-j=\half}, w \in \IZ$ (for the left-movers). 
We show that those topological closed string observables found in \cite{Mukhi} with $j$ outside the unitary range can be written as operators descended from $j=-\half$ in spectrally flowed sectors. 

We then turn to the open string sectors of the two theories. The $c=1$ model has two types of branes -- the first type (ZZ branes \cite{Zamolodchikov}) is localized in the strong coupling region and is parameterized by two integers $(m,n)$. Out of these, only the first in the series, the $(1,1)$ brane, is unitary in the Lorentzian theory. The second  type (FZZT branes \cite{Fateev, Teschner}) extends to infinity towards the weak coupling end of the Liouville direction, and dissolves at a  finite value of the Liouville coordinate. These branes are parameterized by a continuous complex parameter $\mu_{B}$ whose absolute value measures the position where the brane dissolves.

The cigar theory has branes which are zero, one and two dimensional. The unitary B-branes which we consider are of two kinds: the $D0$-branes, with no continuous parameter, which are localized near the tip of the cigar and carry a charge under the RR zero form\footnote{Note that we will be considering always the Euclidean theory on the cigar and the ``$D0$-brane''  is really an instanton.}; and the $D2$-brane, filling the cigar, that are parameterized by a continuous parameter $\t \mu_{B}$ which also measures the extent of the brane towards the tip. We show that the one-point functions for both the localized/extended branes when restricted to the topological sector of the coset agree respectively with those of the ZZ/FZZT branes of the $c=1$ string.

We then proceed to compare the open string chiral primaries on the cigar branes using the annulus amplitude, and find that they indeed match the physical open string excitations of the $c=1$ branes.
The $D0$ brane in the topological theory has two physical open string states that match the two perturbative physical degrees of freedom of the ZZ branes in the $c=1$ string theory at self dual radius. The $D2$ brane in the coset has a countably infinite set of chiral primary open string excitations (similar to the closed string case). We compute the two point functions of these open chiral primaries and find that they match the correlators of physical operators on the FZZT branes in the $c=1$ string theory.

\subsubsection*{Organization}

In section \ref{susycigar}, we review the supercoset theory and its twisting, in order to establish conventions for the sequel.  The $A$-twist of this model is equivalent to the $c=1$ theory at self-dual radius.
In section \ref{closedmap}, we map the closed string operators by analyzing the null vectors in the Verma modules of $SL(2,\IR)$ and using the Wakimoto representation of $SL(2,\IR)$. We also identify, using spectral flow, the operators in the Hilbert space of the parent $SL(2,\IR)$ theory which (after descent to the coset) map to the $c=1$ states. 
In section \ref{open}, we show that the one-point functions defining the localized and extended D-branes are equivalent in the two theories, and match the open string marginal deformations of the branes as well as their reflection amplitudes. In appendix \ref{vertop}, we collect a few useful facts about vertex operators on the cigar. In appendix \ref{extended}, we exhibit an extended $\CN=2$ algebra in the asymptotic variables of the cigar.

\section{The twisted $SL(2,\IR) / U(1)$ coset} \label{susycigar}

\subsection{The $\CN=2$ structure}
Let us first review the coset construction of Kazama-Suzuki \cite{Kazama} that leads to an $\CN=2$ superconformal algebra. We follow the conventions in \cite{janbranes} which are natural from the point of view of the cigar geometry and perform the axial gauging of the $U(1)$ subgroup of the $SL(2,\IR)$. The $\CN=1$ currents of the parent $SL(2,\IR)_k$ theory has currents ($J^a, \psi^a)$ that satisfy the OPE
\begin{align}
J^a(z)\, J^b(w) &\sim \frac{g^{ab}\, k/2}{(z-w)^2}+\frac{f^{ab}_c\, J^c}{z-w} 
\cr
J^a(z)\, \psi^b (w) &\sim \frac{i f^{ab}_c\, \psi^c}{z-w} 
\cr
\psi^a(z)\,\psi^b(w) &\sim \frac{g^{ab}}{z-w} \,.
\end{align}
Our conventions are such that $g_{ab} = \diag(+,+,-)$ and $f^{123} = \bar{f}^{123} = 1$.
In order to define the $\CN=2$ currents, it is convenient to first define
\be
j^a = J^a - \hat{J}^a = J^a + \frac{i}{2}f^a_{bc}\,\psi^b\,\psi_c 
\,.
\ee
The currents ($j^a$, $\jb^a$) commute with the free fermions $(\psi^a,\psib^a)$ and generate a bosonic $SL(2,\IR)$ at level $k+2$. The Hilbert space of the original $\CN=1$ $SL(2,\IR)_k$ model therefore factorizes into a purely bosonic $SL(2,\IR)_{k+2}$ and $3$ free fermions. We now implement the quotient of the action of $(J^{3}, \psi^{3})$ following Kazama and Suzuki \cite{Kazama}. The $\CN=2$ currents are given by:
\begin{align}\label{nequalstwo}
T &= T_{SL(2,\IR)} - T_{U(1)} 
\cr
G^{\pm} &= \sqrt{\frac{2}{k}}\, \psi^{\pm}\,j^{\mp} 
\cr
J^R &= \frac{2}{k}\,j^3+\left(1+\frac{2}{k}\right)\,\hat{J^3} 
=  \frac{2}{k}\, J^3 + \psi^+\, \psi^- 
\end{align}
where the $U(1)$ energy momentum tensor is 
\begin{align}
T_{U(1)} &= -\frac{1}{k}\, J^3 J^3 + \frac{1}{2}\psi^3\p \psi^3  \,,
\end{align}
and where we have used the formulae $\psi^{\pm} = \frac{\psi^1\pm i\psi^2}{\sqrt{2}}$ and $j^{\pm} = \frac{j^1\pm i j^2}{\sqrt{2}}$. These conventions are chosen so that the superscript denotes the charge. Note that we have conventions that are completely left-right symmetric and so we have not specified the right movers explicitly. The currents in \eqref{nequalstwo} generate an $\CN=2$ superconformal algebra with central charge $c=3+\frac{6}{k}$.

\subsection{Gauging}\label{gauging}
To clarify the geometry and construct the operator spectrum of the coset, it is useful to follow the procedure \cite{Dijkgraaf} of gauging the $U(1)$ symmetry by the addition of a gauge field. The axial gauging of the coset is done by adding an extra boson $X$, and restricting to cohomology of  an additional BRST charge, whose currents are given by 
\begin{align}
J_{BRST} &= C\,(J^3+ i\sqrt{\frac{k}{2}}\, \p X) + \gamma'\,(\psi^3 + \psi^X)  \,,
\end{align}
and similarly for the right-movers. Here, $(B,C)$ is a $(1,0)$ ghost system
associated with this BRST symmetry with central charge $c=-2$. 
The fields $X,\Xb$ are the left and right moving components of the boson which will be identified with the angular direction of the cigar geometry. The $\b',\gamma'$ superghosts remove the contributions of the fermions $\psi^3$ and $\psi^X$ leaving the two free free fermions on the cigar $\psi^{\pm}$. The gauging currents are defined to be
\footnote{A Note on Conventions :
We can denote a group element of $SL(2,\IR)$ as \cite{Maldacena}
$g = e^{i\frac{(t+\varphi)}{2}\sigma_2}\, e^{\rho \sigma_3} \, e^{i\frac{(t-\varphi)}{2}\sigma_2}$.
Here, $\rho$, $t$ and $\phi$ are the global coordinates on $SL(2,\IR)$. The action of the left and right currents $J^3$ and $\Jb^3$ on $g$ is of the form $g \longrightarrow h_L \cdot g \cdot h_R^{-1}$.
Thus, the generator $J^3+\Jb^3$ generates translations around the compact $\varphi$
direction, which results in the quantization of the $J^3+\Jb^3$
eigenvalue. Furthermore we treat $X+\Xb$ as the geometric coordinate and
according to \eqref{gaugecurrent}, we identify it with $\varphi$. Thus, we
gauge
axially the non-compact direction $t$, as in \cite{Dijkgraaf} (though
the conventions are different). The closed string background therefore is identical to the euclidean black hole described in \cite{Witten}.}
\be\label{gaugecurrent}
J_g = J^3 + i\sqrt{\frac{k}{2}}\p X  \qquad \Jb_g = \Jb^3 + i \sqrt{\frac{k}{2}}\bar{\p}\, \Xb \,.
\ee
From the definition, it is easy to check that it is a ``null-current" and has
non-singular OPE's with $T$, $J^R$ and also with itself. 

\subsection{Wakimoto representation of $SL(2,\IR)$}
In what follows, we will make extensive use of the Wakmoto free field representation of the $SL(2,\IR)$ currents in terms of which
\begin{align}
j^+ &= \beta \quad&  \jb^+ &= -\bar{\beta}\, \bar{\gamma}^2 - \sqrt{2k}\, \bar{\gamma}\, \bar{\p}\phi - (k+2)\, \bar{\p}\bar{\gamma}\cr
j^3 &= - \beta\,\gamma - \sqrt{\frac{k}{2}}\, \p\phi \quad& \jb^3 &= \bar{\beta}\,\bar{\gamma} + \sqrt{\frac{k}{2}}\, \bar{\p}\phi  \cr
j^- &= \beta\, \gamma^2+ \sqrt{2k}\, \gamma\, \p\phi + (k+2)\, \p\gamma \quad& \jb^- &= -\bar{\beta} 
\end{align}
Note that we have opposite sign conventions on the right moving sector. We will see that this is convenient to perform the $A$-twist. The energy momentum tensor in these variables is given by
\begin{align}
T_{SL(2,\IR)} &= \beta\,\p \gamma - \half (\p\phi)^2 -\frac{1}{\sqrt{2k}}\, \p^2\phi -\half\psi^+\p\psi^--\half\psi^-\p\psi^+\,.
\end{align}
We collect the various fields and some of their properties  in Table \ref{untwistedtable}.
\begin{table}
\begin{center}
\begin{tabular}{|c|c|c|c|c|}
\hline
 Field & $\Delta$ & Q & c \\
\hline\hline
X & $-$  & $0$ & $1$  \\
\hline
$\phi$ & $-$ & $\sqrt{\frac{2}{k}}=\sqrt{2}$ & $1+\frac{6}{k}=7$ \\
\hline
$(\psi^+,\psi^-)$ & $(1/2,1/2)$ & $-$ & $1$ \\
\hline
$(\psib^+,\psib^-)$ & $(1/2,1/2)$ & $-$ & $ 1$ \\
\hline
$(\b,\gamma)$ &$(1,0)$ & $-$ & $2$ \\
\hline
$(B,C)$ & $(1,0)$ & $-$ & $-2$ \\
\hline
 \end{tabular}
\end{center}
\caption{List of fields in the untwisted theory, their conformal weight $\Delta$, the background charge $Q$ for the bosons and the central charge $c$.\label{untwistedtable}}
\end{table}

\subsection{Twisting}\label{twisting}
From this section on, we restrict attention to the case $k=1$, i.e. to an
$N=2$ superconformal theory at central charge $c=9$. 
However in subsequent sections, we retain the symbol $k$ for the level to clarify and interpret many of the operations to be discussed. The twisting in the more general case at arbitrary level $k$ has been discussed in \cite{Takayanagi}. We will perform the $A$-twist in what follows.
It is defined by modifying the energy momentum tensor as
\be\label{twist}
T \longrightarrow T + \half \p J^R \qquad \bar{T} \longrightarrow \bar{T} - \half \p \Jb^R \,,
\ee
where $J^R$, $\bar{J}^R$ are the left and right $R$-currents that appear in \eqref{nequalstwo}.
Using the explicit expressions for the currents in terms of the Wakimoto free fields, the twisted energy momentum tensor can now be written as
\begin{align}
T &= - \p \b \gamma - \half (\p\phi)^2 - \sqrt{2}\, \p^2\phi - \half(\p X)^2 - \half\psi^+\p\psi^- -\half \psi^-\p\psi^+  +\frac{3}{2}\p(\psi^+\psi^-) \cr
\bar{T}& = -\p \bar{\b} \bar{\gamma} -\half (\bar{\p}\phi)^2 - \sqrt{2}\, \bar{\p}^2\phi -\half(\bar{\p} X)^2 - \half\psib^+\p\psib^- -\half \psib^-\p\psib^+ -\frac{3}{2} \bar{\p} (\psib^+ \psib^-) \,.
\end{align}
We collect in Table \ref{twistedtable}, the conformal dimensions and central charges of the twisted theory.
\begin{table}[h]
\begin{center}
\begin{tabular}{|c|c|c|c|}
\hline
 Field & $\Delta$ & Q & c \\
\hline\hline
X & $-$  & $\sqrt{2k}-\sqrt{\frac{2}{k}}=0$ & $1$  \\
\hline
$\phi$ & $-$ & $\sqrt{\frac{2}{k}} + \sqrt{2k} = 2\sqrt{2}$ & $1+\frac{6(k+1)^2}{k}$ = 25 \\
\hline
$(\psi^+,\psi^-)$ & $(-1,2)$ & $-$ & $ - 26$ \\
\hline
$(\psib^+,\psib^-)$ & $(2,-1)$ & $-$ & $ - 26$ \\
\hline
$(\b,\gamma)$ &$(0,1)$ & $-$ & $2$ \\
\hline
$(B,C)$ & $(1,0)$ & $-$ & $-2$ \\
\hline
 \end{tabular}
\end{center}
\caption{List of fields in the twisted theory, their conformal weight $\Delta$, the background charge $Q$ for the bosons and the central charge $c$.\label{twistedtable}}
\end{table}
We can easily recognize the field content of the bosonic $c=1$ string by identifying $X$ with the matter field and the field $\phi$ with the Liouville field \cite{Mukhi, Ohta1, Ohta2}. The fermions on the cigar get identified with the $(b,c)$ ghosts related to reparametrization invariance on the worldsheet. But note from Table \ref{twistedtable} that it is $\psi^-$ and $\psib^+$ that become spin-$2$ anti-ghosts on the left and right sectors. The BRST currents are therefore given by
\be\label{topBRST}
Q_{top} := Q^+ = \oint G^+ = \oint \psi^+ j^- \quad \hbox{and} \quad \bar{Q}_{top} := \bar{Q}^- = \oint \Gb^- = \oint \psib^-\, \jb^+ \,,
\ee
while the other twisted supercurrents are identified with
\be\label{bghosts}
G^- = \psi^-\, \b = b\,\beta \qquad \hbox{and} \qquad \Gb^+ = \psib^+\, \bar{\b} =\bar{b}\, \bar{\b}\,,
\ee
where we have used the Wakimoto representation for the $j^+$ and $\jb^-$ currents. The quartet of $(\b,\gamma)$ and $(B,C)$ have total central charge zero. As shown in the appendix of \cite{Mukhi}, after restricting to the cohomology of the two BRST charges associated with the gauging and twisting, they decouple from the rest of the fields leaving behind the pure $c=1$ bosonic theory.

\section{Mapping of closed string cohomology}\label{closedmap}

\subsection{Observables of the topological cigar}\label{cprimary}
In this section, we derive the map of closed string operators in the two
theories along the lines of \cite{Mukhi}. In the next section, we will interpret most of the
operators as being descendants of operators in spectrally flowed sectors. But
in the present subsection, we allow the parameter that labels the Casimir of
the discrete representations, $j$,  to take all half-integer values. 
It is well known that the observables of the topological $A$-model \cite{BCOV}
are the chiral primaries ($(c,a)$ operators) in the NS-NS sector
of the untwisted theory. These states are annihilated by $G^+_{-1/2}$ and $\overline{G}^{-}
_{-1/2}$ which implies the relation $2\Delta = Q$ and $2\overline{\Delta}= -\bar{Q}$ between the 
conformal dimension and R charge on the left and right sector.

\subsubsection{Verma modules of the $SL(2,\IR)_3$ Kac-Moody algebra}
\begin{figure}[h]
\centering
\includegraphics[]{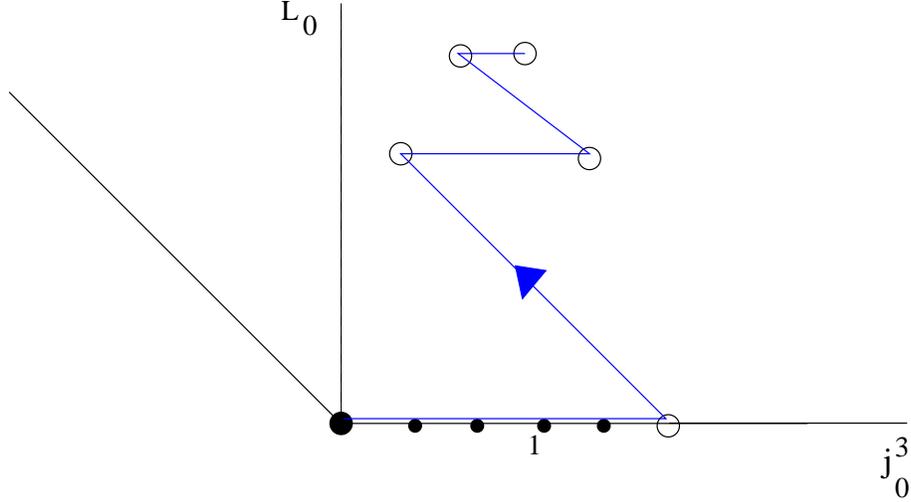}
\caption{
The weight diagram for a Verma module built on a lowest weight representation $D^+_{-j}$ of
$SL(2,\IR)$ with Casimir parametrized
by $j \ge -\half$, with primary 
marked by the big full dot, and null vectors marked by open white circles. 
(The small black dots on the horizontal axis are calibrations.) \label{nullplpos}}
\end{figure}

\begin{figure}[h]
\centering
\includegraphics[]{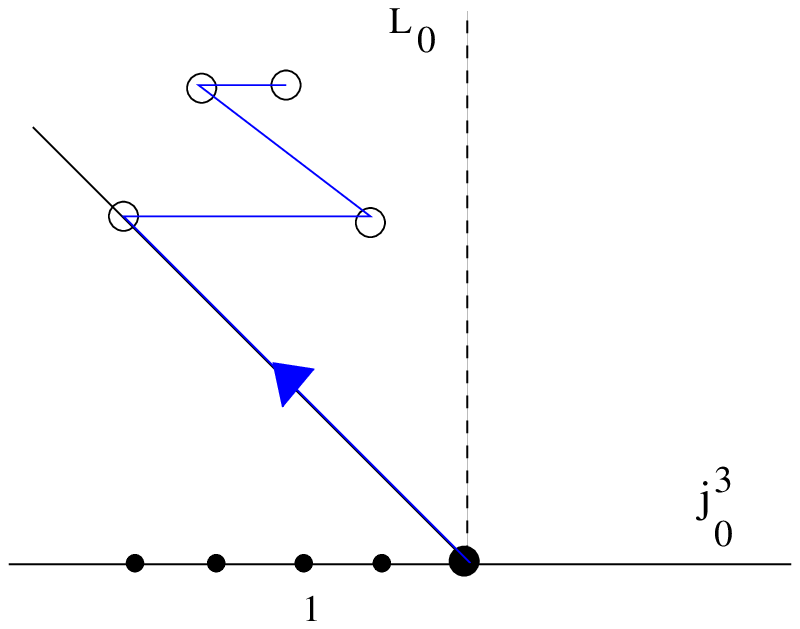}
\caption{
The weight diagram for a Verma module built on a lowest weight representation $D^+_{-j}$  with Casimir   $j \le -\half $, with primary marked by the big full dot, and null vectors marked by open white circles.\label{nullplneg}}
\end{figure}

We shall restrict ourselves to the left-moving sector of the theory and we consider states in 
theory that belong to the Wakimoto modules $W_j$. A most useful result we will use is the 
following \cite{Frenkel} : 
\begin{itemize}
\item For $j \ge -\half$, the Wakimoto modules $W_j$ are isomorphic to the Verma modules $V_j$
 built on the lowest weight state (LWS) representation $D^+_{-j}$, while for $j \le -\half$, 
the Wakimoto modules are isomorphic to the dual Verma modules $V^*_{j}$ built on the dual lowest 
weight space $(D^+_{-j})^*$.
\end{itemize}
A lowest weight state is a chiral primary state, as it is annihilated by $j^-$ (and hence by 
$G^+$, from \eqref{nequalstwo}) and has $m_{bos}=-j$. The Verma module built on this state 
has null states which in turn are 
also lowest weight states. Each of these states will give rise to a chiral primary in the 
cigar and hence to an observable of the topological coset\footnote{For a general proof of this 
statement, see \cite{Ashok2}.}. We exhibit, in figure \ref{nullplpos} and figure \ref{nullplneg}, 
the location of null vectors in the Verma modules $V_{j}$ for $j \ge -\half$ and $j \le -\half$ 
respectively.

The location of the null states in the weight diagram of the Verma module can be obtained from
 the Kac-Kazhdan determinant formula \cite{Kac}. For $D^+_{-j}$, there is a null state for 
every pair of integers $(r,s)$ which satisfy
\be
2j+ 1 = r+s \quad\hbox{such that} \quad r\cdot s > 0 \quad \hbox{or} \quad r > 0 \quad\hbox{and}\quad s=0 \,.
\ee 
The dimension and $j^3$ eigenvalue of the null vector relative to the original LWS are 
respectively given by $\Delta h= r\cdot s$ and $\Delta m_{bos} = r$. Let us consider the cases
 $2j+1 \le 0$. There are two exceptional cases : 
\begin{list}{$\bullet$}{\setlength{\leftmargin}{0cm}\setlength{\itemindent}{0.5cm}}
\item For $2j+1=0$, we see from the Kac-Kazhdan formula that the Verma module is irreducible. 
The only primary is the original LWS state with $j=-m_{bos}= -\half$.
\item For $2j+1= -1$, or $j=-1$, there are again no null states among the descendants and the module is irreducible. 
\end{list}
For $2j+1 \le -2$, we always have a null vector with $s=-1$ and $r=2j+2$. The Verma module 
built on this null vector is reducible and its null vectors coincide with all other null 
vectors present in the original LWS module. On the other hand, for a Verma module with $2j+1 
> 0$, there is a null vector with $s=0$ and $r=2j+1$ and again, the Verma module built upon 
this null vector includes all other null vectors. Using these two facts, we find that the 
original LWS module is reducible and can be resolved into a string of irreducible modules 
\cite{Marcus, Mukhi}. For $j \ge -\half$, we find (for integer $j$) 
\be
I_j \longrightarrow I_{-j-1} \longrightarrow I_{j-1} \longrightarrow \ldots \longrightarrow I_0 \longrightarrow I_{-1}
\ee
where $I_{j}$ refers to the irreducible representation built on a LWS with $m_{bos}= -j$. Similarly for $j \le -1$, we get (for integer $j$)
\be
I_j \longleftarrow I_{-j-2}  \longleftarrow I_{j+1} \longleftarrow \ldots \longleftarrow I_0 \longleftarrow I_{-1}
\ee
We summarize these results by labeling the $j^3_0$ spectrum of lowest weight states as
$S^{+}_{>}=\{-j,\ldots ,j+1 \}$ for $j \ge -\half$ and $S^{+}_{<} = \{-|j|+2,\ldots , |j|\}$ for $j \le -1$.

%\begin{figure}[h]
%\centering
%\includegraphics[]{nullmipos}
%\caption{
%The weight diagram for a Verma module built on a lowest weight representation $D^-_{j}$  with 
%Casimir   $j \ge 0$, with primary 
%marked by the big full dot, and null vectors marked by open white circles.
%}
%\end{figure}
%\begin{figure}[h]
%\centering
%\includegraphics[]{nullmineg}
%\caption{
%The weight diagram for a Verma module built on a lowest weight representation $D^-_{j}$  with 
%Casimir   $j \le -\half $, with primary 
%marked by the big full dot, and null vectors marked by open white circles.}
%\end{figure}

\subsubsection{Cohomology of $Q$}\label{qcohom}
For every such null vector, we will find a state in the $c=1$ BRST cohomology as follows : 
from each of the lowest weight primary states, we first construct representatives of the
 BRST cohomology of the topological cigar theory. In order to do that, we consider the BRST 
operator: 
\be
Q= Q_{top} + Q_{U(1)} = \int  \left[ c j^-  + C \left(j^3 +cb+ \frac{i}{\sqrt{2}} \partial X \right) \right]\,.
\ee
In the cohomology relative to the zero-mode of $C$, we consider only states annihilated by 
\be
j^3_0+ \int \frac{i}{\sqrt{2}} \partial X + \int cb \,,
\ee 
which leads to the equation for states in the relative cohomology : 
\be\label{shiftm}
-p_X=m= (m_{bos}+n_{gh}) \,.
\ee
where $n_{gh}$ is the ghost number. 

Now, we can  construct states in the BRST cohomology of the topological cigar theory by 
considering the following states in the Wakimoto modules $W_j$ (see \cite{Polchinski} for conventions for ghost sector): 
\be\label{abstractstate} 
| \downarrow \rangle \otimes | LWS \rangle 
\ee
combined with the state with the appropriate $X$-momentum fixed by \eqref{shiftm}. These states are at ghost number $-1/2$. The corresponding operator, by the state-operator correspondence \cite{Polchinski} is at ghost number one. Thus, we find operators corresponding to states in the BRST cohomology which are at ghost number one. These operators, given the spectra for $m_{bos}$ found above for the highest weight states, and the link between the $m$-quantum number, $m_{bos}$ and the ghost charge of the state \eqref{shiftm}, have a spectrum of $m$-values which is $S^+_>=\{ -j-1/2, \dots ,j+1/2 \}$ and $S^+_<=\{-|j|+3/2,-|j|+1/2,\dots, |j|-1/2 \}$. Notice that the latter spectrum can be split as $S^+_<=\{ |j|-1/2 \} \cup \{-|j|+3/2,\dots, |j|-3/2 \}$. This will prove useful in identifying the $c=1$ states to which these chiral primaries are mapped.

We first consider the case $j \ge -\half$. Defining $s=j+\half$,  the range of
$X$-momentum in $S^+_{>}$ fills out a spin-$s$ $SU(2)$-representation. It will
be convenient to provide free field (Wakimoto) representations of certain
cohomology elements at this point. We choose the sign of $\phi$-momentum to be opposite to the sign of $j$ (modulo a constant shift). Demanding that the $SU(2)$-highest weight state with ($p_X = j+\half$) has dimension zero in the twisted theory, we find 
\begin{align}\label{Vtachyon}
\CV_j &= c \, e^{-\sqrt{2} \,\phi} \, e^{ \sqrt{2} \left( j+\half \right)(-\phi + i X)}  = c e^{-\sqrt{2}(s+1)\phi} \, e^{i\sqrt{2}sX}\equiv Y^-_{s,s}\,.
\end{align}
In the last equality, we have identified the vertex operator with the tachyonic
state at ghost number one denoted $Y^-_{s,s}$ in \cite{Witten}. The rest of
the null vectors in $S^+_{>}$ are obtained by acting alternately with
$j^+_{0}$ and $j^-_{-1}$  in the $SL(2,\IR)$ representation. This leads to
operators with the same $\phi$-momentum but with $X$-momentum decreased by
one. This operation is equivalent to acting with the coset analog of the
$SU(2)$ lowering operator  $K^-$ (introduced in \cite{Mukhi}) on the operator in formula \eqref{Vtachyon} 
\be\label{kminus}
K^-  = \oint\, \beta\, e^{-i\sqrt{2}X} \,.
\ee
In this way we can show that the elements in the topological coset cohomology which arise from
 all the lowest weight states in the Verma module with $j \ge -\half$ correspond to the 
elements $Y^-_{s,n}$ of the $c=1$ bosonic string cohomology (row one of Table \ref{qcoh}).

We turn to the case $j \le -1$, the elements in $S^+_{<}$, which has the opposite sign for the Liouville momentum. Again, the lowest weight states gives rise to elements in the cohomology of ghost number $1$. The state with $m = - p_X = |j|-\half$ can once again be given a Wakimoto representation. Defining $s = -(j+\half)$, we get the state with the lowest $X$-momentum in the $SU(2)$ $s$-multiplet :
\be\label{yplus}
Y^+_{s,-s} = c \, e^{\sqrt{2}(s-1)\phi} \,e^{- i\sqrt{2} sX}\,.
\ee
The operator corresponding to the next null vector in the weight diagram is found by 
acting with $(j^-_{-1})^{2 |j|-1}$. The resulting operator is
\be\label{Bj}
\CB_{j}\vert_{j=-s-\half}= c\, \gamma^{2 s-1}\, e^{\sqrt{2} (s-1) \phi}\, e^{ i \sqrt{2} (s-1) X} \,.
\ee  
We claim that this operator is identical to the operator $a\cO_{s-1,s-1}$, which is a discrete state at ghost number one, which is the partner of a ground ring element $\cO_{s-1,s-1}$. The notation for the elements of the cohomology is taken from \cite{WittenZ}, where $a$ is defined to be  
\be\label{definea}
a = [Q,\frac{\phi}{\sqrt{2}}] \,.
\ee
This is an operator in the $Q$-cohomology as it is $Q$-closed and not $Q$-exact if we restrict to the space of conformal fields (since $\phi$ is not a conformal field). Using the explicit Wakimoto representation of the coset theory, we get \cite{Mukhi}, 
\be
a = c\gamma \,.
\ee
Substituting this in \eqref{Bj}, we see that if our identification is correct, then the ground ring element $\cO_{s-1,s-1}$ has a Wakimoto form
\begin{align}
\cO_{s-1,s-1} &= \gamma^{2 (s-1)}\, e^{\sqrt{2} (s-1) \phi}\, e^{ i \sqrt{2} (s-1) X} \cr
& = \left [\gamma^2 \, e^{\sqrt{2} (\phi+i X)} \right]^{s-1} = (\cO_{\half,\half})^{2(s-1)}\,.
\end{align}
The $\cO$ are the discrete states at ghost number zero which form a ring on account of regular
 OPEs with themselves. We see that we have recovered the ring structure obtained in 
\cite{Witten} thereby confirming our identification. Note that using our definition of $a$ in 
the coset, $\cO_{0,0}=1$ as expected. We also observe that $a=\CB_{-3/2}$ and $a \in 
D^+_{3/2}$. Once the top element $\cO_{s-1,s-1}$ has been obtained, the lowering operator 
$K^-$ can be used to construct all the ghost number zero operators $\cO_{s-1,n}$ and their 
partners $a\cO_{s-1,n}$. Formally, the operation of removing $a$ from an operator is defined 
by acting with the operator $G^-_{-1/2}$. 

To summarize, we see that the elements in the topological coset cohomology which arise from 
all the lowest weight states in the Verma module of $D^+_{-j}$ with $j \le -1$ correspond to 
the ghost number $1$ elements $\{Y^+_{s,-s}\,, aO_{s-1,n} \}$ of the $c=1$ bosonic string 
cohomology (row three
 of Table \ref{qcoh}). To check that there are the right number of operators $Y^-$ relative to
 the number of operators $a O$, it suffices to count their number at a {\em given} value of 
the quadratic Casimir for the $D^+_{-j}$ representation. One finds then the expected relative 
number of operators. 

%Next, we observe that also from the highest weight states, we can construct non-trivial elements in the BRST cohomology. Since the states now need to be of the form $\ket{\uparrow} \otimes \ket{HWS} = c_0 \ket{\downarrow } \otimes \ket{ HWS}$ (to be annihilated by the BRST charge), the spectrum of the bosonic $j^3_0$ quantum numbers is now shifted oppositely to find the $m$-spectrum, namely by $+1/2$. This also implies that the operators we find now are at ghost number two, by the state-operator correspondence. The operator now contains a factor $c \partial c$, at ghost number two. 

%The spectrum of $m$ values is given by $S^{-}_{>}=\{-j-\half,\ldots ,j+\half \}$ for $j \ge -\half$ and $S^-_{<} = \{-|j|+\half,\ldots , |j|-\frac{3}{2}\} = \{-|j|+\half\} \cup \{-|j|+\frac{3}{2},\ldots , |j|-\frac{3}{2}\}$ for $j \le -1$. We can repeat the analysis performed for the $D^+_{-j}$ representations. Once again, the sign of $j$ determines the sign of the Liouville momentum. We find that for $j\ge -\half$, these operators correspond to the ghost number two operators $aY^+_{s,n}$ on the $c=1$ side while for $j\le -1$, we get the operators $P_{s-1,n}$ while the state with $m=-p_X=-|j|+\half$ gives rise to the operator $Y^-_{s,s}$. (Again one can check the relative number of these operators at given quadratic Casimir.) 

\begin{table}[h]
\begin{center}
\begin{tabular}{|c|c|c|c|c|}
\hline
Repn & $n_{gh}$ & s & Spectrum & $c=1$ Operator  \\
\hline\hline
$D^+_{-j}\,$ $j \ge -\half$ & $1$ & $j+\half$ & $ -j-\half, \dots ,j+\half $ & $Y^-_{s,n}$ \\
\hline
 & $2$ & $j+\half$ & $ -j-\half, \dots , j-\half$ & $aY^-_{s,n}$ \\
\hline
$(D^+_{-j})^*\,$ $j \le -\half$ & $1$  & $-\left(j+\half\right)$ & $-|j|+\frac{3}{2},\dots, |j|-\half$
 & $a\CO_{s-1,n}\,,\, Y^+_{s,-s}$ \\
\hline
& $0$ & $ -\left(j+\half \right)$ & $-|j|+\frac{3}{2},\dots, |j|-\frac{3}{2} $ & $\CO_{s-1,n}$
\\
\hline
 \end{tabular}
\end{center}
\caption{Partial List of Operators in $Q$-Cohomology\label{qcoh} }
\end{table}
As we have already seen in the case of the ground ring elements, once we have these basic 
operators, we use the techniques in \cite{WittenZ} to
generate partners of the existing cohomology elements by either multiplying by
$a$ or by taking away a factor of $a$. This generates partner states at ghost
numbers one greater or one less than the state one started off with. The inverse $a$-operation
 is carried out by using the operator $G^-_{1/2}$ in the coset theory. In the twisted theory 
and in the Wakimoto variables, this coincides with the operation $\beta b_0$ (in agreement 
with \cite{WittenZ}). We summarize our results in Table \ref{qcoh} and mention that these 
operators are in one to one correspondence with the rather abstract list of cohomology 
elements at ghost number $0$,$1$ and $2$, in the Appendix of \cite{Mukhi} by E. Frenkel. 

We observe that the operators in Table \ref{qcoh} account for half of all the BRST
cohomology elements in the $c=1$ string, as enumerated in \cite{WittenZ}. The rest of the 
cohomology elements are the BPZ duals of the ones listed in table \ref{qcoh} and with whom the
operators in table \ref{qcoh} have non-trivial two point functions. The BPZ duals are in one 
to one correspondence with the $\{Y^+_{s,n}\}, \{P_{u,n}, aY^+_{s,n} \}$ and $\{aP_{u,n}\} $ 
at ghost numbers $1$, $2$ and $3$ respectively. This completes the first analysis of the map 
between the closed string observables of the twisted coset and the $c=1$ string.

%\begin{figure}[h]
%\centering
%\includegraphics[scale=0.5]{wittenzdiag}
%\caption{The states in the $c=1$ string theory. We denote $u=s-1$. Ghost
%  number increases from zero to three from left to right. The boxed elements
%  at ghost number one $Y^-$ and $a\cO$ originate from $D^+_{-j}$
%  representations  in the coset while the ghost number two states 
%$aY^+$ and $P$ originate from the $D^-_j$ representations. The 
%tachyon operators $Y^{\pm}_{s,\pm s}$ are not shown in the diagram.  }
%\end{figure}

\subsection{$c=1$ observables from spectrally flowed representations}\label{specflow}

As mentioned in the introduction, the full spectrum of the $SL(2,\IR)_{1}$ theory within the improved unitarity bound  \cite{Maldacena, Pakman} is given by $\sum_{w} D^{+,w}_{1/2}$. The states of the coset therefore have to descend from this Hilbert space. It is therefore important to check whether the states in the cohomology of $Q_{top}$ we have found (that match to the $c=1$ string) can be rewritten as states descending from spectrally flowed representations. In this section, we show that this is indeed the case. We first make a few observations about spectral flow:
\begin{list}{$\bullet$}{\setlength{\leftmargin}{0cm}\setlength{\itemindent}{0.5cm}}
\item Consider a state in the coset $\ket{s1}$ with
\be
\ket{s1}\, :\,  L_{0}^{cig}=-j(j+1)+m^{2}\quad J^{3}_{0}=m\quad \hat{J^{3}_{0}} = 0 \quad J^{R}_{0}=2m \,.
\ee
This descends from a state in the parent theory  with 
\be
L_{0}=-j(j+1)\quad J^{3}_{0}=m \quad \hat{J^{3}_{0}} = 0 \,.
\ee
A spectral flow by $w$ units takes this state to another state with
\be
L_{0}=-j(j+1)+ \frac{mw}{ 2} -\frac{w^{2}}{ 4} \quad J^{3}_{0}=m+\frac{w}{ 2} \quad \hat{J^{3}_{0}} = -w \,.
\ee 
This state descends in the cigar to $\ket{s2}$ with 
\be
\ket{s2}\, : \, L_{0}^{cig}=-j(j+1)+m^{2} \quad J^{3}_{0}=m+\frac{w}{ 2} \quad \hat{J^{3}_{0}} = -w \quad J^{R}_{0}= 2m \,.
\ee 
In particular, the dimension and $R$-charge of the state in the coset does not change. The fermion number ( $=$ ghost number in the $c=1$ theory) changes by $w$ units.

\item
The operator $G^{-}_{1/2}$ in the untwisted theory changes the dimension and $R$-charge of a state by $(-\half, -1)$, and therefore acts on the $Q_{top}$ cohomology. This operator also decreases the fermion number by $1$ unit. Recall that this is identical to the inverse-$a$ operation mentioned in subsection \ref{qcohom}. 
\end{list}

Following the construction in section \ref{cprimary}, we now give an algorithm that allows us to obtain all physical states of the $c=1$ theory starting with the operator $\Phi_{-\half,\half}$ and doing the operations of spectral flow, finding descendants using the operators $j^+_0$ and $j^-_{-1}$ and the ``inverse-a" operation (acting with $G^-_{1/2}$). All these are operations strictly within the untwisted coset theory. The resulting states in the twisted theory will be BRST cohomology elements. 
\begin{list}{$\bullet$}{\setlength{\leftmargin}{0cm}\setlength{\itemindent}{0.5cm}}
\item
Start with the operator $\Phi_{-\half, \half}$ which has dimension and $R$-charge given by $2L^{cig}_{0}=J^{R}=1$ in the untwisted theory. From Table \ref{qcoh}, we see that in the topological theory, this maps to the state 
\be
\Phi_{-\half,\half}= Y^+_{0,0} = Y^-_{0,0} = c\, e^{-\sqrt{2}\phi} \,. 
\ee
We will obtain other states in the $c=1$ theory by repeatedly applying a two-step procedure : (a) spectral flow by one unit, (b) act by $G^{-}_{1/2}$. The operator that implements the spectral flow can be explicitly written out in Wakimoto variables as\footnote{
This operator was introduced in \cite{Mukhi}.} 
\be
U = c\, \b^{-1}\, e^{\frac{i}{\sqrt{2}}X}\, e^{-\frac{\phi}{\sqrt{2}}} \,.
\ee
Acting with $U$ on $Y^{\pm}_{0,0}$ results in 
\be
c\, \p c\, \beta^{-1}\, e^{-\frac{3}{\sqrt{2}} \phi}\, e^{\frac{i}{\sqrt{2}}X} = aY^-_{\half,\half}\,,
\ee 
where the explicit representatives are given in \cite{Mukhi}. From Table \ref{qcoh}, this operation (in coset variables) takes a state $Y^{-}_{0,0} \in D^+_{1/2}$ to a state $aY^-_{\half,\half} \in D^-_{-1}$ and amounts to exchanging the two limits of the improved bound $-1/2 \ge j \ge -1$. This confirms that $U$ indeed implements spectral flow. Acting on this state with $G^-_{1/2}$ removes the factor of $a$ and we thus get $Y^-_{\half,\half}$. The corresponding operator in the untwisted theory has $2L^{cig}_{0}=J^{R}=0$. 

More generally, if we start with a state at ghost number one with $2L^{cig}_{0}=J^{R}=-j$, what we have shown is that spectral flow by one unit $w=-1$ preserves the $2L^{cig}_{0}=J^{R}$ as shown above. Then act by $G^{-}_{1/2}$ to get a state at ghost number one with $2L^{cig}_{0}=J^{R}=-j-1$. In this way, one can generate all the states of the form $Y^-_{s,s}$. Once we have generated the chiral states with arbitrary low dimension, we can use the construction of subsection \ref{qcohom} and get the rest of $a Y^-_{s,n}$ -- acting with $j^{+}_{0}, j^{-}_{-1}$. 

\item
Let us now spectral flow in the opposite direction (using $U^{-1}$) starting with the same operator $Y^{\pm}_{0,0}$, but now reversing the steps : (a) multiply by $a$, (b) act with $U^{-1}$. We get 
\be
c\, e^{-\frac{i}{\sqrt{2}}X}\, e^{-\frac{1}{\sqrt{2}} \phi} = Y^+_{\half,-\half} \,.
\ee 
where we have again taken explicit representatives from \cite{Mukhi}. Repeating this two step procedure we generate all states of the form $Y^+_{s,-s}$ (and their ghost number two partners). As before, we can use the construction of \ref{qcohom} and get the rest of $Y^+_{s,n}$ by acting with $j^{+}_{0}, j^{-}_{-1}$. We have thus completely generated the states in the $2$nd and $4$th rows of Table \ref{qcoh} as follows :
\be
\ldots\xleftarrow{a}Y^+_{\half,-\half} \xleftarrow{U^{-1}} aY^+_{0,0} \xleftarrow{a} Y^{\pm}_{0,0} \xrightarrow{U} aY^-_{\half,\half} \xrightarrow{G^-_{1/2}}Y^-_{\half,\half}\xrightarrow{U}\ldots 
\ee

\item
But now, note from Table \ref{qcoh} that the elements $Y^+_{s,-s}$ and $Y^-_{s,s}$ {\it also} appear in the $3$rd and $1$st rows respectively. Thus, the descent operations on these particular states (i.e. finding null vectors in the Verma modules) will give us all the discrete states at ghost number $1$, as well as their partners at ghost number two/zero obtained by multiplying/inverting $a$. Thus, we can reconstruct all the elements in Table \ref{qcoh}.
\end{list} 
 
To summarize, what we have shown is that given the operator $\Phi_{-\half,\half} \in 
D^+_{\half}$ along with the states one can obtain from it by spectral flow, and using 
operations intrinsic to the coset theory (such as acting with $G^-_{1/2}$, $j^-_{0}$ and 
$j^+_{-1}$), one can generate {\it all} physical operators in the $c=1$ theory. We would like 
to point out that most of these chiral operators in the coset have $J^{R}_{0} \neq 1$. The 
only operators which do are the chiral primary \CN=2 Liouville interaction operators, which 
map to the Liouville interaction in the $c=1$ theory.

\subsection{Relative normalization}
So far, the analysis of section \ref{qcohom} has enabled us to identify the quantum numbers of the operators on the coset side that maps to the tachyonic operators $\CV_j$ in the $c=1$ string and in equation \eqref{Vtachyon}, we have an explicit realization in terms of Wakimoto variables. However, in the next sections where we compare disc one point functions that define D-branes in the two theories, the relative normalization of the closed string fields becomes relevant. 

In this section, we fix the precise relative normalization between the cigar chiral primaries and the $c=1$ tachyonic operators $\CV_j$ by using the reflection amplitudes for closed strings in the two backgrounds. In particular, we will consider certain winding operators on the $c=1$ string side that will couple to Neumann branes (in the $c=1$ direction). We recall that there are operators in the $c=1$ theory which are the reflections of \eqref{Vtachyon} off the Liouville wall:
\be\label{Vttachyon}
\t\CV_{j} = c\, \cb\, e^{-\sqrt{2}\left(j+\frac{3}{2}\right) \phi} \, e^{-i\sqrt{2}\left(j+\half \right) (X-\Xb)} \,.
\ee
These operators\footnote{In \cite{Mukhi}, these were obtained by acting on
  $\CV_j$ with $j^+_0$ $2j+1$ times. This leads to powers of $\b$
  which can be put to a constant consistently since $j^+= \sqrt{\mu}$ is a spin $0$ operator.} have the opposite $X$ momentum as $\CV_j$, and the same exponent of $\phi$. They are related to the operators \eqref{Vtachyon} as \cite{Fateev} (with $\gamma(x)=\frac{\Gamma(x)}{\Gamma(1-x)}$):
\begin{align}\label{relation}
\CV_j & = S(j)\,\t\CV_{-j-1}\, \cr
S(j) & = -(\pi\mu\gamma(1))^{-2j-1}\,\left(\frac{\Gamma(2j+2)}{\Gamma(-2j)}\right)^2 \,.
\end{align}
From the cigar theory, we can simply understand them as reflections off the tip. Operators in the coset with $j\ge -\half$ and $j\le -\half$ are also related by a reflection amplitude  \cite{Ponsot, Ribault} (we are restricting to specific $(c,a)$ fields):
\be\label{cosetref}
\Phi_{j,-j,j}=R(j,-j,j)\, \Phi_{-j-1,-j,j} \,.
\ee
with
\begin{align}
R(j,-j,j) &= -(\t \mu\gamma(1))^{-1-2j}\frac{\Gamma(-j+m)\Gamma(-j-\mb)(\Gamma(2j+1))^2}
{\Gamma(1+j+m)\Gamma(1+j-m)(\Gamma(-2j-1))^2} \cr
& \stackrel{m= -j}{\longrightarrow} \left[ - (\t \mu\gamma(1))^{-1-2j}
\left(\frac{\Gamma(2+2j)}{\Gamma(-2j)}\right)^2\right] \cdot (\Gamma(-2j))^2 \,.
\end{align}
Since the observables $\cV_j$ in the $c=1$ string are mapped to $\Phi_{j,-j}$ of the coset, we see from \eqref{cosetref} that $\t\CV_{j}$ maps to the operators $\Phi_{j,j+1}$ in the coset. We shall use these amplitudes to  fix the relative normalization between the coset and $c=1$ string operators. We may write
\be\label{norm}
\Phi_{j,-j,j} = \cN^2 \, S(j)\,\Phi_{-j-1,-j,j} \qquad \hbox{with} \qquad \CN = \Gamma(-2j) \,,
\ee
which tells us that
\be\label{normop}
\Phi_{j,-j,j} \mapsto \Gamma(-2j)\, \CV_{j} \quad \hbox{and} \quad \Phi_{-j-1,-j,j} \mapsto \frac{1}{\Gamma(-2j)}\,  \t \CV_{-j-1} \,.
\ee
 We will see later that these relative normalizations are indeed the natural ones when we compare the disc one-point functions of the respective branes in the two theories.

\subsection{Some connections to the conifold geometry}
In this section, we briefly discuss a few target space geometrical aspects of the bulk theory.
We note that this subsection is technically somewhat orthogonal to 
the rest of the paper. 
We discuss some features of the well known claim \cite{Ghoshal} that the $c=1$ string theory is equivalent to the topological B-model on the deformed conifold defined as a hypersurface in  $\IC^4$ :
\be\label{defconifold}
\det\left(\begin{array}{cc} a_1 & a_4 \cr a_3 & a_2 \end{array} \right) = a_1a_2 -a_3a_4 = \mu \,.
\ee
To support the claim, we review in some detail the construction of the analog
of the holomorphic three-form in the $c=1$ bosonic string language. The holomorphic three-form can locally be written as:
\be
\Omega = \frac{da_1 da_2 da_3}{a_3}.
\ee
We can translate the holomorphic three-form in terms of the operators that generate the ground ring using the relations $a_1=xx', a_2=yy', a_3= xy', a_4=yx'$, where
\begin{align}
x & = O_{\half, \half} \qquad y = O_{\half,-\half} \cr
x' &= \bar{O}_{\half, \half} \qquad y' = \bar{O}_{\half, -\half} \,.
 \end{align}
The resulting expression for the holomorphic threeform can be compactly written as:
\be\label{Omega}
\Omega = \omega \wedge \theta' - \theta \wedge \omega'
\ee
 where we have made use of the following definitions for the volume form $\omega$ and
 the one-form $\theta$ on the $(x,y)$-plane:
 \be
 \theta = x dy - y dx \qquad\hbox{and} \qquad \omega = d \theta = dx \wedge dy,
 \ee
and similarly for the right-movers. Using the duality between forms and vector fields \cite{WittenZ}, one can dualize the three-form to the vector field $S$:
 \be\label{S} 
 S = x \partial_x + y \partial_y - x' \partial_{x'} - y' \partial_{y'}.
 \ee
In \cite{WittenZ}, there is a precise dictionary between physical vertex operators in the $c=1$ string and vector fields and differential forms on the conifold. From the above reasoning, 
the vertex operator that corresponds to the holomorphic $3$-form can now be read off 
to be 
\be
\Omega \mapsto a+\bar{a},
\ee
where $a$ is the ghost number one operator introduced in \eqref{definea}. The vector field $S$ in \eqref{S} measures the difference between left and right-moving Liouville momentum  in the $c=1$ bosonic string theory. All functions on the quadric defined by \eqref{defconifold}  must be $S$-invariant. Indeed this is equivalent to the statement that all operators in the $c=1$ theory have equal left-right Liouville momentum \cite{WittenZ}. 

There are an infinite number of left and right moving currents on the worldsheet $J_{s,n,n'}$ which are generated from the elements of the cohomology using a descent procedure described in \cite{WittenZ}. The currents that generate the left and right $SU(2)$ action are those with $s=1$. These are symmetries of the $c=1$ string and so symmetries of the $B$-model on the deformed conifold. In the geometric description, these symmetries are therefore generated by vector fields whose Lie derivative action on the holomorphic $3$-form vanishes
$\cL\cdot\Omega = 0$. The explicit expressions for these vector fields can be obtained from the discussion in the appendix of \cite{Witten}. The $s=1$ currents are given by\footnote{These coincide with the expressions in equation $(2.24)$ of \cite{Witten} when expressed in the $(x,y,x'y')$ variables.}
\begin{align}
X &= a_1 \frac{\p}{\p a_4} + a_4 \frac{\p}{\p a_2} \qquad Y = - a_4\frac{\p}{\p a_1} - a_2\frac{\p}{\p a_3} \cr
Z &= a_4 \frac{\p}{\p a_4} + a_2 \frac{\p}{\p a_2} - a_1 \frac{\p}{\p a_1} - a_3 \frac{\p}{\p a_3}
\end{align}
They satisfy the algebra
\be
[X,Y] = Z \qquad [Z,X] = -X \qquad \hbox{and} \qquad [Z,Y] = Y \,.
\ee
These vector fields generate the left $SU(2)$ currents and one can check that the Lie derivative of each of these vector fields leaves $\Omega$ in \eqref{Omega} invariant. The simplest way to do this is to define
\be
\Lambda = df \wedge \Omega = da_1\,da_2\, da_3\, da_4\,,
\ee
and show that $\cL_{X} \Lambda = \cL_{X} df = 0$ and similarly for $Y$ and
$Z$. That concludes our review of some features and symmetries of the conifold
geometry and its connection to the $c=1$ string language. We move on from the
closed string to the open string sector in the next section.

\section{ D-branes and open strings}\label{open}
In this section, we shall present the boundary states for the $B$-branes in the topological theory on the cigar by starting with the known BPS boundary state on the physical cigar theory, and restricting to the chiral primaries. We shall show that the one-point functions of these operators precisely match the one-point functions of the $c=1$ closed string operators (described in the previous section) on the branes when the matter part of the brane is at the point corresponding to Neumann boundary conditions. Further, by analyzing the vacuum annulus diagram and disk diagrams in the topological theory, we show that the open string spectrum and boundary two point functions are the same.

\subsection{The branes of the $c=1$ theory}
In the $c=1$ theory,  the D-branes are described by  a boundary state in the CFT describing the free boson at self-dual radius tensored with a boundary state in the Liouville theory at $c=25$
 with one BRST condition imposed.

In the $c=1$ matter sector, the boundary states are those of $SU(2)_{1}$ which span \cite{Recknagel} the three-dimensional $SU(2)$ group manifold. The boundary condition breaks the $SU(2)_{L} \times SU(2)_{R}$ of the closed strings to the diagonal $SU(2)_{+}$ under which the open string states are classified.

As described in the introduction, the Liouville theory admits two types of boundary conditions. one corresponding to branes localized in the strong coupling region and the other extended in the Liouville direction to the weak coupling region.
The  localized ZZ branes do not have any further parameters, whereas the extended FZZT branes have  an additional modulus given by the complex parameter $\sigma$ which is related to the position in the Liouville direction where the brane dissolves.

In the open string sector, the ZZ branes admit only the identity
representation, so the only allowed marginal operators 
are the three Goldstone bosons corresponding to the broken generators. At the north pole on the 3-sphere\footnote{In the circle description, this point corresponds to Neumann boundary conditions. The operators $e^{\pm i \sqrt{2}X}$ become marginal only at the self dual radius.}, these are $(i c \p X, c e^{\pm i \sqrt{2}X})$.  The operator $c \p X$ however is null in this theory, as can be seen from the negative sign in the partition function of the ZZ brane in the $c=1$ theory\footnote{This can be seen by tensoring the ZZ partition function \cite{Zamolodchikov} with the annulus amplitude of the $c=1$ Neumann brane \cite{Recknagel}}. When $X$ has the interpretation of time, this operator is the gauge field $A_{0}$ which does not have a kinetic term. There can be however physical Wilson lines of this gauge field when $X$ is a Euclidean compact direction (as we consider).

The FZZT branes on the other hand admit in the open string sector all the continuous and discrete representations. The physical operators at ghost number one  \cite{GhoshalMM} are labeled as $V^{l}_{n}$ where $(l,n)$ are the quantum numbers of the conserved $SU(2)$, $2l \in {\bf Z}, -l \le n \le l$.
The $(l,l)$ operators are explicitly given by\footnote{In contrast to the closed strings, the even dimensional representations of $SU(2)$ are not allowed. we have kept the convention $\alpha'_{open}=  \alpha'_{closed}$ in writing the above vertex operators on the branes. This introduces a factor of two in the vertex operators \eqref{openvop} as a function of $l$ compared to the closed strings.}
\be\label{openvop}
V^{bdry}_{l} = c e^{\rt2 (2l+1) (i X - \varphi) - \rt2 \varphi}.
\ee

The 3 modes of dimension 1 on the FZZT branes correspond to the three Goldstone modes of the broken generators. Turning on these modes changes the value of the zero mode corresponding to the broken symmetry.

\subsection{D0/ZZ branes}
\subsubsection{Boundary state}
The $(1,1)$ ZZ brane in the $c=1$ string theory is a tensor product of two boundary states: the $(1,1)$ ZZ brane in Liouville theory and the Neumann brane in the matter sector. The one point function of a closed string operator therefore factorizes as
\be
\vev{\CV_{j}} = \vev{e^{-\sqrt{2}(j+\frac{3}{ 2}) \phi}} \cdot \vev{e^{i\sqrt{2}(j+\half) (X-\Xb)}} \,.
\ee
The disc one-point functions for the ZZ branes are normalized so that $\vev{1}=1$. This is useful as all momentum independent factors drop out of the expressions for the one-point function. The non-trivial piece of the one point function arises from the Liouville part of the operator \eqref{Vtachyon} and can be read off from \cite{Zamolodchikov} to be (with $b=1$ and $Q = 2$ in their conventions)
\be\label{ZZonepoint}
\vev{V_j}  = 2\cdot (\pi \mu \gamma(1))^{-j-\frac{3}{2}}\cdot \frac{1}{\Gamma(-2j-1) \Gamma(-2j)} \, .
\ee

In the coset theory, the properly normalized one-point function is \cite{Eguchi, Ribault, janbranes, Ahn, hosomichi}:
\begin{align}\label{janonepoint}
 \frac{\Psi^{D0}(j\,; \, -j,-j)}{\Psi^{D0}(-\frac{3}{2}\, ; \, \frac{3}{2},\frac{3}{2}) } &= (\t \mu\gamma(1))^{-\half-j} \frac{\Gamma(-2j)\Gamma(0)}{\Gamma(-2j-1)\Gamma(-2j)}\cdot (\pi\mu\gamma(1))^{-1}\,
 \frac{\Gamma(1)\Gamma(3)}{\Gamma(3)\Gamma(0)} \cr
&=\left[\frac{\Gamma(-2j)}{\Gamma(3)}\right] \cdot 2\ (\t \mu\gamma(1))^{-j-\frac{3}{2}}\frac{1}{\Gamma(-2j-1)\Gamma(-2j)}
\end{align}
Taking into account the normalization factor \eqref{normop} for the two vertex operators with $j=j$ and $j=-\frac{3}{2}$, we see that there is exact match with the $c=1$ one point function.

\subsubsection{Open string spectrum}
The open  string spectrum on the topological branes can be deduced from the annulus amplitude by  restricting to the open string chiral primaries. The localized branes have only the identity representation in the open string channel e.g. \cite{janbranes}, and the NS sector open string partition function ($k=1$) is given by
\begin{align}\label{NSpartition}
Z^{NS}_{1,1}(\tau,\nu) &= \frac{\Theta_3(\tau,\nu)}{\eta(\tau)^3}\sum_{s\in \IZ + \half} \frac{1}{1+yq^s}\left(q^{s^2-s}y^{2s-1}-q^{s^2+s}y^{2s+1}\right) \cr
& = \frac{\Theta_3(\tau,\nu)}{\eta(\tau)^3}\sum_{s\in \IZ + \half} \left(q^{s^2-s}y^{2s-1}(1-yq^{s})\right)\,.
\end{align}
The total number of chiral primaries in the open channel can be obtained simply by writing out the Ramond sector partition function and counting the number of Ramond ground states. This can be easily done in our case. The result is
\be\label{openR}
Z^{R}_{1,1}(\tau,0) = 2+4q+12q^2+\ldots
\ee
Thus, we infer that there are two chiral primary states. The dimension and R-charge of these states can be found by now expanding \eqref{NSpartition} as
\begin{align}
{\rm Tr_{NS}}\; q^{L_0} y^{J_{0}} & =  q^{\frac{c}{ 24}} Z^{NS}_{1,1}(\tau,\nu) \cr
& = 1+q+(y^{-3}+y^{-1}+y+y^3) q^{\frac{3}{2}} + (y^{-2}+3+y^2)q^2+ \dots \cr
\end{align}
It is easy to see that the two chiral primary states have dimension and R-charge $(\Delta,Q^R)$ given by $(0,0)$ and $(\frac{3}{2},3)$. The first state with $(j=m=0)$ and its reflected operator\footnote{As in the closed string case, operators with $j+j'=-1$ are reflected into each other, and the topological theory will only count ({\it e.g.} from the Ramond sector ground states \eqref{openR}) one for each such pair of operators.}  match the two physical open string states $e^{\pm i \rt2 X}$  on the ZZ branes in the $c=1$ theory discussed earlier.  Note that this is simply the chiral part of the closed string map \eqref{Vtachyon}, \eqref{Vttachyon}, after being careful about normalizations. We make a few remarks about the spectrum : 
\begin{list}{$\bullet$}{\setlength{\leftmargin}{0cm}\setlength{\itemindent}{0.5cm}}
\item The $(j=-m=0)$ state maps to the {\it spin $\half$ representation} in the closed string theory \eqref{Vtachyon}, \eqref{Vttachyon}.  In the open string theory, it maps to operators in the {\it spin $1$ representation}. The reason is the same as to why only odd dimensional reps are allowed on the FZZT branes in this theory \cite{GhoshalMM}, and boils down to the fact that the conformal dimension (R charge) of a closed string operator is the sum of the left and right dimensions (R charge). Operationally, this is the factor of two in \eqref{openvop}.
\item The other state with $(-j=m=\frac{3}{ 2})$ and its reflected counterpart corresponds to the ghost number zero states $\CB_{j}, \t \CB_{j}$ described in \eqref{Bj} and discussed in \cite{Mukhi, Takayanagi}. 
\item The operator corresponding to the state $c \p X$ is not present in the spectrum, as on the ZZ branes of the $c=1$ theory. There is however one other mode which is not a physical perturbative open string state which enters the definition of the B-brane boundary state as the zero mode of the scalar corresponding to the R-current. This corresponds to the Wilson line on the ZZ brane.
\item The two open string ``moduli''  above are not dimension one operators on the physical branes in the superstring theory involving this coset, such a brane would have no moduli ({\it e.g.} the localized branes constructed in \cite{Fotopoulos:2005cn, Ashok}).
\end{list}

\subsection{D2/FZZT branes}
Let us introduce a bit of notation to facilitate the matching of boundary states between the extended D2 branes of the coset and the FZZT branes in $c=1$ string theory. The FZZT branes are semiclassically defined by a boundary interaction of the form \cite{Fateev}
\be
S_{bdry} = \mu_B\, \int_{\p\Sigma}\,  e^{\phi} dx \,,
\ee
where $x$ is a coordinate along the boundary. The one point functions are solutions to certain recursion relations which are obtained by looking at a bulk two point function involving specific (bulk) degenerate fields. The result for the one point functions are written in terms of a uniformization parameter $s$ related to $\mu_B$ as $\cosh^2(\pi b s) = \frac{\mu_B^2}{\mu} \sin\pi b^2$ where we have to take the limit $b\rightarrow 1$. Here $\mu$ is the bulk cosmological constant. We shall explicitly write them out in the following section. For now, we focus on a similar discussion of extended branes in the coset CFT following \cite{hosomichi}.

The discussion in \cite{hosomichi} is in the context of $\CN=2$ liouville theory with boundary, which is just the mirror theory to the one we have been considering. In particular, there is a precise map between the extended A-branes in that theory and the extended B-branes we consider. These can again be described in terms of a boundary interaction term
\be
S_{bdy} = -\t \mu_B\, \int dx (\lambda\bar{\lambda}  - \bar{\lambda}\lambda)(\psi^+\psi^- - i\sqrt{2} \p\theta) e^{- Q \phi}
\ee
The common link between the two boundary interactions is that these are ``holomorphic square roots" of bulk screening operators \cite{hosomichi}. The $\lambda$ are boundary fermions that introduce Chan-Paton factors on the boundary. The one point functions are again computed using techniques similar to the Liouville case and are specified by a complex parameter $\sigma$ which is related to $\t \mu_B$ above as $\t\mu_B = \frac{1}{2\pi k}\,(\t\mu \gamma(1))^{\half}\Gamma(-\frac{1}{k})\cos(\frac{\pi}{k}\sigma) $. 

\subsubsection{Boundary state}
Let us compare the one point functions of FZZT branes and the extended $D2$ branes of the supersymmetric coset theory. The non-trivial piece arises from the Liouville factor, and can be read off from \cite{Fateev} to be (putting $b=1$)
\be
\vev{V_j} = (\pi \mu \gamma(1))^{-j-\half} \Gamma(2j+1)\Gamma(2j+2) \cosh(\pi s(2j+1))\,.
\ee
Note that this is the unnormalized one point function and we have only kept the $j$-dependent pieces. The relevant wavefunction in the coset theory can be obtained from \cite{janbranes,Ribault} to be
\begin{align}
\Psi^{D2}(j\,;\,-j,-j) & = (\t\mu \gamma(1))^{-j-\half} \left[\Gamma(-2j) e^{i\sigma(2j+1)}+\frac{\Gamma(0)}{\Gamma(2j+1)} e^{-i\sigma(2j+1)} \right] \times \cr
& \qquad \qquad  \times \Gamma(2j+1)\,\Gamma(2j+2) \,. \cr
& = \Gamma(-2j) \cdot 2\,(\t\mu \gamma(1))^{-j-\half }\,\Gamma(2j+1)\,\Gamma(2j+2)\, \cos(\pi \sigma(2j+1) ) \,.
\end{align}
Thus, taking into account the normalization \eqref{normop}, we see that the unnormalized one point functions once again match if we identify the FZZT parameter ($s$) with the semi-classical $B$-field on the $D2$ brane ($\sigma$) as $is=\,\s$.

\subsubsection{Open string spectrum and two point function}
The open string spectrum is deduced from the annulus amplitude \cite{janbranes} as before. The open string spectrum on extended $D2$-branes on the cigar  is independent of $\mu_{B}$ and the boundary state allows all the continuous and discrete representations in the open string channel. The analysis of the spectrum therefore reduces to that of the (chiral part of) the closed string spectrum as in section \ref{closedmap}. Restricting attention to the chiral primaries, we find that the topological open string operators labeled by an half-integer $l$ can be mapped to the marginal operators $V^{bdry}_{l}$ of the FZZT branes.

Note that in this case, like the closed strings, but in contrast to the localized branes, the three operators corresponding to motion along the broken $SU(2)$ directions are all physical.

The boundary operators have a boundary fermion number, and each brane thus has a two dimensional Chan-Paton index. The open string operators have even or odd boundary fermion numbers at each end and are thus two by two matrices. This fermion number maps to the ghost number as for the closed strings, and while comparing boundary two point functions, we restrict ourselves to physical operators at fermion number $1$.

The boundary open string reflection amplitude on the FZZT branes is:
\begin{multline}
d(l|s_{1}, s_{2})  =
  \nu_{FZZ}^{j+\half} \frac{{\bf G}(-2l-1)}{{\bf G}(2l+1)} \; \times \cr
 \frac{1}{{\bf S}(l+\frac{3}{2}+i \frac{s_{1}+s_{2}}{2}) {\bf S}(l+\frac{3}{2}-i \frac{s_1-s_2}{2}){\bf S}(l+\frac{3}{2}+i \frac{s_1-s_2}{2}){\bf S}(l+\frac{3}{2}-i \frac{s_1-s_2}{2})}
\end{multline}
where
$\nu_{FZZ}=\pi \mu \gamma(b=1)$.
The boundary two point function for chiral operators with fermion number one ($l+m=0$) on the $D2$-brane on the cigar is \cite{hosomichi}:
\begin{multline}
d^{-1}_{\lambda}(l,m|J_{1}, J_{2})  =  \frac{\G(-l+m)}{\G(l+m+1)}  \nu^{l+\half} \frac{{\bf G}(-2l-1)}{{\bf G}(2l+1)} \; \times \cr
\frac{1}{{\bf S}(l+J_{1}+J_{2}+\frac{5}{ 2}) {\bf S}(l+J_{1}-J_{2}+\frac{3}{ 2}){\bf S}(l-J_{1}+J_{2}+\frac{3}{ 2}){\bf S}(l-J_{1}-J_{2}+\half)}\cr
= \Gamma(-2l) \cdot \nu^{j+\half} \frac{{\bf G}(-2l-1)}{{\bf G}(2l+1)} \; \times \cr
\frac{1}{{\bf S}(l+\frac{3}{2}+i \frac{s_{1}+s_{2}}{ 2}) {\bf S}(l+\frac{3}{2}+i \frac{s_{1}-s_{2}}{ 2}){\bf S}(l+\frac{3}{2}-i \frac{s_{1}-s_{2}}{ 2}){\bf S}(l+\frac{3}{2}-i \frac{s_{1}+s_{2}}{ 2})}
\end{multline}
where the parameter $J$ is related to $\sigma$ as $\sigma = 2J+1$. This makes it clear that we need to identify $is= \sigma$, just as with the one-point functions. One can observe that with this identification, these two expressions agree upto a normalization factor which we observe to be the square root of the  factor in \eqref{norm}, \eqref{normop} as expected for a boundary operator.

\section{Closing Remarks}
We have revisited the correspondence between the topological cigar
at level one and the $c=1$ string theory and clarified the discussion in \cite{Mukhi} regarding the mapping of bulk operators. We have reproduced the states listed in the appendix of \cite{Mukhi} in our analysis, and in our description the identification of the states in the list as null vectors in the Verma modules of the affine $SL(2,\IR)$ current algebra is made manifest.\footnote{We believe that our analysis may clarify properties of the $c=1$  cohomology used in the comparison of the $c=1$ string amplitudes to the amplitudes derived from the Landau-Ginzburg model as described in \cite{Hanany:1994ya}.} The same procedure also allowed us to precisely map the states in the list to specific representatives of the topological observables using the Wakimoto variables. 

We also showed that the states in the (twisted) super-coset to which the states of the $c=1$ string are mapped, descend from the Hilbert space of the parent $SL(2,\IR)_{1}$ theory derived in  \cite{Maldacena}, $\sum_w D^{+w}_{-j=1/2}$. The states that naively seem to descend from representations labeled by different $j$ values are understood to descend from the spectrally flowed representations.

These results teach us a little about the relation of the topological 
theory on the cigar with the physical superstring theory. To compare the spectrum and 
correlation functions in the $A$ model on the cigar to those of the $c=1$ theory, we use the 
untwisted cigar SCFT and restrict to states which formally obey the chiral primary condition. 
Most of these states are not of dimension one, 
but could be tensored with other states to produce on-shell states 
in a  superstring theory on a larger spacetime involving the cigar, as done in 
\cite{Argurio:2000tb,Rastelli}.

We matched the one point functions that define the D$0$/D$2$ branes in the topological coset with the ZZ/FZZT branes of the $c=1$ string theory. Further, we mapped the open string deformations for the two kinds of branes and also the boundary two point function on the extended D$2$ and FZZT brane. 
The statements in the above paragraph also apply to the open string modes on the branes in the topological theory.

At various places in the paper, we have a discussion of the $SU(2)$ symmetry intrinsic to the $c=1$ theory. The closed $c=1$ string theory has a $SU(2) \times SU(2)$ symmetry, and this is also a symmetry of the topological string theory on the coset. However, the currents that generate these symmetries are not  dimension one currents
in the untwisted SCFT and and so do not generate symmetries of the superstring theory involving this coset ({\it e.g.} cigar $\times \IR^{1,3}$).

Both the ZZ and the FZZT branes (and their topological counterparts) preserve a diagonal $SU(2)$.
The  $D2$/FZZT branes have three marginal operators $(e^{\pm i \rt 2X}, \p X)$ which correspond to motion in the directions of the broken generators. On the $D0$/ZZ brane, one of these operators $(\p X)$ becomes null, and only its zero mode survives. Based on the identification of the closed string theory with the topological theory on the deformed conifold, it is natural to identify the localized (compact) brane with the Lagrangian brane wrapping the $S^{3}$ of the deformed conifold.

We believe that our analysis could be useful in recovering the BPS sector of the gauge theory physics for branes on the conifold from an exact worldsheet description. It would be especially interesting to apply similar techniques to more general non-compact Calabi-Yau manifolds. For instance, it would be instructive to compute the superpotential for scalars arising from combinations of D-branes wrapping cycles in the Calabi-Yau purely from the worldsheet theory.

Our analysis could also be helpful to better understand the topological 
subsector of the AdS/CFT correspondence, and more generally, holography. 
Recently, it was shown \cite{Rastelli} that the Liouville field emerges as a holographic 
direction in topological theories involving the $SL(2)$ current algebra. 
In the last couple of years, holography in theories involving the Liouville theory 
has been understood better using D-branes in these theories \cite{McGreevy, Klebanov:2003km}. 
The identification of the D-branes in $c=1$ and the D-branes in 
the topological cigar goes a step towards completing this circle of ideas.

We close by noting that the techniques used to map the closed string 
observables are not specific to the cigar at level one. In particular, one can
do a similar computation for the $c < 1$ theories coupled to Liouville
theory. Recently it was suggested in \cite{Takayanagi} that these are related
to $c <1$ non-minimal matter coupled to Liouville. It would be interesting to repeat
the null vector analysis for $SL(2,\IR)$ at general level $k$ to get a better grip
on the cohomology of the twisted supercosets, i.e. of the topological string
on these $N=2$ superconformal backgrounds, from first principles.

\section*{Acknowledgements}

We are grateful to Bobby Acharya, Freddy Cachazo, Eleonora Dell'Aquila, Emanuel Diaconescu, Edi Gava, K. S. Narain, Vasilis Niarchos, Christian R\"omelsberger, Martijn Wijnholt and especially Jaume Gomis for very helpful discussions. S.A. would like to thank the ASICTP and its members for their hospitality where this work was initiated. J.T. would like to thank the PI and its members for their hospitality where part of this work was carried out jointly. Research of S.A at the Perimeter Institute is supported in part by funds from NSERC of Canada and by MEDT of Ontario. The work of the S.M. and J.T. is partially supported by the RTN European program: MRTN-CT2004-503369.

\begin{appendix}  
 \section{Vertex operators on the cigar} \label{vertop}
In this appendix, we construct the primaries of the supersymmetric coset $SL(2,\IR)/U(1)$ at level $k$. We follow the conventions of \cite{Giveon, janbranes} and references therein. The primaries of the bosonic $SL(2,\IR)_{k+2}$ are denoted $V^{j}_{m_{bos}\mb_{bos}}$, where $m_{bos}$ is the charge under the purely bosonic $j^3$ current
\be
j^3(z)V^{j}_{m_{bos}\mb_{bos}}(0) \sim \frac{m_{bos}V^{j}_{m_{bos}\mb_{bos}}}{ z} \qquad  \jb^3(\zb)V^{j}_{m_{bos}\mb_{bos}}(0) \sim \frac{\mb_{bos}V^{j}_{m_{bos}\mb_{bos}}}{ \zb}\,.
\ee
They have (left and right) conformal dimensions
\be
\Delta(V_{j,m_{bos},\mb_{bos}}) = \overline{\Delta}(V_{j,m_{bos},\mb_{bos}}) =  -\frac{j(j+1)}{ k} \,.
\ee
As these fields are independent of the free fermions, they are also primary fields of the superconformal $SL(2,\IR)$ at level $k$. In order to obtain the primaries of the coset, it is useful to bosonize the various currents as follows:
\begin{align}\label{leftscalars}
i\p H &=  \psi^{+}\psi^{-}  \qquad  J^3 = -\sqrt{\frac{k}{2}}\p X_3 \cr
J^R &= i\sqrt{\frac{c}{3}}\p X_R  \qquad  j^3 = -\sqrt{\frac{k+2}{2}}\p x_3 \,,
\end{align}
where the normalizations ensure that the scalars have canonical OPEs. Since we have left-right symmetric conventions, there are similar expressions for the right-moving sector as well.
These scalars are not all independent and using the definition of the bosonic currents and the $\CN=2$ algebra \eqref{nequalstwo}, we can rewrite all scalars in terms of $X_3$ and $X_R$:
\begin{align}\label{Hexpansion}
\sqrt{\frac{k}{2}} x_3 = -i X_R + \sqrt{\frac{k+2}{2}} X_3 &\qquad \hbox{and} \qquad iH = -\sqrt{\frac{2}{k}} X_3 + i \sqrt{\frac{k+2}{k}} X_R   \,.
\end{align}
The right moving currents have a similar expansion in terms of scalars. Given these expressions, and knowing that the current that is gauged in the coset are $j^3$ (in the bosonic case), we can decompose
\begin{align}\label{bosonicdecomp}
V^j_{m_{bos},\mb_{bos}} &= \Phi^j_{m_{bos},\mb_{bos}} e^{\sqrt{\frac{2}{k+2}}(m_{bos}x_3+\mb_{bos}\xb_3)}\cr
&  \equiv \Phi^j_{m_{bos},\mb_{bos}}
e^{\frac{2i }{\sqrt{(k+2)k}}(m_{bos}X_R+\mb_{bos}\Xb_R)}\, e^{\sqrt{\frac{2}{k}}(m_{bos}X_3+\mb_{bos}\Xb_3)}
\end{align}
where $\Phi^j_{m_{bos},\mb_{bos}}$ is a primary of the bosonic Euclidean coset CFT (at level $k+2$). One infers that
\begin{align}
\Delta(\Phi^j_{m_{bos},\mb_{bos}}) &=  -\frac{j(j+1)}{k} + \frac{m_{bos}^2}{k+2}\cr
\overline{\Delta}(\Phi^j_{m_{bos},\mb_{bos}}) &= -\frac{j(j+1)}{k} + \frac{\mb_{bos}^2}{k+2}
\end{align}
In the supersymmetric coset, we also gauge the fermionic current $\psi^3$ and the primary we start with in the parent theory is of the form $V^j_{m_{bos},\mb_{bos}}\, e^{inH+i\nb \Hb}$. The coset primaries are obtained by using \eqref{bosonicdecomp}\ and the expression for $H$ in \eqref{Hexpansion}.
These two equations lead to the decomposition of the primary in the parent theory of the form (we suppress the right movers)
\begin{align}\label{fermionicdecomp}
V^j_{m_{bos}}\,e^{inH} &= \Phi^{j}_{m_{bos}}\, e^{i\left(\frac{2m_{bos}}{k+2}+n\right)\sqrt{\frac{k+2}{k}}X_R}\, e^{\sqrt{\frac{2}{k}}\,(m_{bos}+n)X_3} \,.
\end{align}
This allows one to infer that the full superconformal coset primary is given by
\be
\Phi^{j,n,\nb}_{m_{bos},\mb_{bos}} = \Phi^j_{m_{bos},\mb_{bos}} e^{i\sqrt{\frac{k+2}{k}}\left[\left(\frac{2m_{bos}}{ k+2}+n\right)X_R + \left(-\frac{2\mb_{bos}}{k+2}+ \nb\right))\Xb_R\right]} \,.
\ee
It follows from equation \eqref{fermionicdecomp}\ that the $J^3,\Jb^3$ eigenvalues of  $\Phi^{j,n,\nb}_{m_{bos},\mb_{bos}}$ are
\be
m = m_{bos}+n \qquad \mb=\mb_{bos}+\nb\,.
\ee
In terms of $(m,\mb)$ and $(n,\nb)$, the left/right conformal dimension is read off to be
\begin{align}\label{dimension}
\Delta(\Phi^{j,n,\nb}_{m_{bos},\mb_{bos}}) &=  -\frac{j(j+1)}{ k} + \frac{m^2}{ k} + \frac{n^2}{ 2} \cr
\overline{\Delta}(\Phi^{j,n,\nb}_{m_{bos},\mb_{bos}}) &= -\frac{j(j+1)}{ k} + \frac{\mb^2}{k} + \frac{\nb^2}{2} \,.
\end{align}
while the $R$-charge is given by
\begin{align}\label{charge}
Q (\Phi^{j,n,\nb}_{m_{bos},\mb_{bos}}) & = \frac{2m}{k} + n\cr
\overline{Q} (\Phi^{j,n,\nb}_{m_{bos},\mb_{bos}}) &= \frac{2\mb}{k} + \nb \,.
\end{align}
In the NSNS sector, we have $n\in \IZ$, while in the RR sector, we have $n\in \IZ+\half$. The axial gauging of the coset is done such that the $J^3$ and $\overline{J}^3$ eigenvalues $m$ and $\overline{m}$ are are related to the asymptotic momentum and winding of the circle direction of the cylinder at infinity \eqref{gaugecurrent}
\be\label{axialgauging}
m = \frac{p+kw}{2} \qquad \overline{m} = \frac{p-kw}{ 2} \,.
\ee

\section{Extended $\CN=2$ current algebra}\label{extended}  % use *-form to suppress numbering

An $N=2$ algebra at central charge $c=9$ allows for an extension with higher spin
currents \cite{Odake:1988bh}. The extended algebra is associative provided the currents
satisfy some extra conditions \cite{Odake:1988bh}. A few explicit realizations of the
associative extended current algebra in terms 
of minimal models are known. 
In this appendix,  we exhibit another realization of this extended current algebra, on the superconformal cigar
at central charge $c=9$. We realize the algebra explicitly in terms of asymptotic variables on the cigar (or
$N=2$ Liouville theory). We then remark on how the operators that extend the algebra map into the $c=1$ 
bosonic string theory.

It was shown in \cite{Odake:1988bh} that a left-moving
 $\CN=2$ superconformal algebra at central charge $c=9$ allows for an extension by two currents of spin $3/2$ and charges $\pm 3$ (which we denote by $R$ and $R^{\ast}$) and two
currents of spin $2$ and charges $\pm 2$ (namely $Y$ and $Y^\ast$).
A constructive way to find this extension is to first consider the operators $R$ and $R^{\ast}$, which 
are (standard) spectral flow operators, and then to close the algebra using the OPE's.
It can be shown that the resulting algebra is associative provided some extra relations between the currents are satisfied \cite{Odake:1988bh}:
\begin{align}
(J^R R)(z) = \partial R (z) \qquad
(J^R Y)+ \frac{1}{2} (G^- R) = \partial Y(z).
\label{associativity}
\end{align}
The algebra in the general case was given in terms of the modes of the currents in \cite{Odake:1988bh}.

We remark in passing that at large volume, the currents $R$ can be constructed using the fact that 
 a unique holomorphic three-form exists on a Calabi-Yau threefold. (For $c=6$,
there exists a similar extension of the current algebra, leading to
a small $N=4$ superconformal algebra. Also for higher values of the central
charge, currents associated to holomorphic forms on the $CY_n$ spaces
can be defined. And, indeed, worldsheet currents can be associated to forms on special holonomy
manifolds, generically.)

In summary, 
we conclude that in the cigar or Liouville conformal field theory at central charge $c=3+\frac{6}{k}$, and at $k=1$, we can enlarge the standard $N=2$ superconformal algebra with the above currents.
\subsection*{In asymptotic variables}
 In the following, we exhibit
explicitly the realization of the $N=2$ superconformal algebra in terms of asymptotic
 coordinates on the cigar (where
in this appendix $Q=\sqrt{2}$). We first recall the $N=2$ superconformal
algebra in these variables (see e.g. \cite{Murthy}):
\begin{align}
T_{as} &= -\frac{1}{2} (\partial \rho \partial \rho+\partial \theta
\partial \theta + \psi_\rho \partial \psi_\rho + \psi_\theta \partial
\psi_\theta + Q \partial^2 \rho) \cr
G^{\pm}_{as} &= \frac{1}{\sqrt{2}} (i (\psi_\rho \pm i \psi_\theta)
(\partial \rho \mp i \partial \theta) + i Q (\partial \psi_\rho \pm i \partial
\psi_\theta) )\cr
J^R_{as} &= - i \psi_\rho \psi_\theta + i Q \partial \theta.
\end{align}
We also define the useful quantities:
\begin{align}
\psi_{cig}^{\pm} &= \frac{1}{\sqrt{2}} (\psi_\rho \pm i \psi_\theta)\cr
-i \psi_\rho \psi_\theta &= \psi_{cig}^+ \psi_{cig}^- = i \partial H.
\end{align}
We now wish to extend this standard $N=2$ algebra and show that it
closes on itself, and that it does satisfy the associativity condition.
We define first of all the left-moving currents which are exponentials in the free
boson corresponding to the $U(1)_R$ current, and which have dimension $3/2$
and charge $3$:
\begin{align}
R &= \sqrt{2} e^{iH+i \sqrt{2} \theta } = \sqrt{2} \psi^+ e^{i \sqrt{2} \theta} \cr
R^\ast &= \sqrt{2} e^{- iH-i \sqrt{2} \theta} = \sqrt{2} \psi^- e^{-i \sqrt{2} \theta}.
\end{align}
%As mentioned above,
%these currents can be interpreted as a left-moving part of the operator
%corresponding to the holomorphic (or anti-holomorphic) three-form
%in the Calabi-Yau three-fold.
Next we record the currents $Y$ and $Y^{\ast}$ that arise from considering the OPE's
of these currents with the supercurrents. These are currents with dimension $2$
and charges $\pm 2$:
\begin{subequations}
\begin{align}
Y &= \frac{i}{\sqrt{2}} e^{i \sqrt{2} \theta}
( \partial \rho + i \partial \theta - i \sqrt{2} \psi_\theta \psi_\rho) \\
Y^\ast &= \frac{i}{\sqrt{2}} e^{-i \sqrt{2} \theta}
( \partial \rho - i \partial \theta + i \sqrt{2} \psi_\theta \psi_\rho).
\end{align}
\end{subequations}
This explicit realization of the algebra can now be used to check straightforwardly
that the equations (\ref{associativity}) for associativity are satisfied.
This gives a new and interesting representation of the 
extended $N=2$ superconformal algebra at $c=9$.

Since we haven't found an encoding of the extended algebra in
terms of OPE's in the literature (although it immediately follows from the
algebra in terms of the oscillators \cite{Odake:1988bh}), we record it here for the readers
convenience :
\begin{subequations}\label{fullOPE}
\begin{align}
G^-(z) R(w)  &\simeq \frac{2 Y(w)}{(z-w)} \\
G^+ (z) R^\ast (w) &\simeq \frac{2 Y^\ast(w)}{(z-w)} \\
G^+(z) Y(w) & \simeq \frac{3 R(w)}{(z-w)^2} +
\frac{ \partial R(w)}{z-w} \\
G^-(z) Y^{\ast} (w) &\simeq \frac{3 R^\ast(w)}{(z-w)^2} +
\frac{ \partial R^\ast(w)}{z-w} \\
R (z) R^\ast(w) & \simeq
\frac{2}{(z-w)^3} + \frac{2 J^R(w)}{(z-w)^2}
+ \frac{(J^R)^2 (w)+ \partial J^R (w)}{(z-w)} \\
R (z) Y^\ast (w) & \simeq \frac{G^+(w)}{(z-w)^2}
+ \frac{(J^R G^+)(w)}{(z-w)} \\
R^\ast (z) Y(w) &\simeq \frac{G^-(w)}{(z-w)^2}
- \frac{(J^R G^-)(w)}{(z-w)} \\
Y(z) Y^\ast(w) & \simeq \frac{3}{(z-w)^4}
+ \frac{2 J^R}{(z-w)^3} + \frac{\half(J^R)^2+T+\partial J^R}{(z-w)^2}
\nonumber \\
&  + \frac{-\frac{1}{2} (G^+G^-)+(TJ)+\frac{1}{4} \partial ((J^R)^2) + \partial T}{(z-w)}.
\end{align}
\end{subequations}
We have checked all these OPE's on the representation in terms of asympotic
variables, and also that the currents $R,R^\ast,Y$ and
$Y^\ast$ have the correct OPE's with the energy-momentum
tensor and the $U(1)_R$ current in accord with their conformal
dimension and R-charge. All other OPE's (except for the standard
$N=2$ superconformal algebra OPE's that are not mentioned above) are regular.

\subsection*{Mapping the algebra onto the $c=1$ string}
We can embed this well-known symmetry structure in the
standard realization of supersymmetric string theory, into the
much larger symmetry group of the 
topological string realization of a subsector of the theory.
To that end, we proceed to translate the currents constructed above into the $c=1$ language. 
The map between the basic twisted $\CN=2$ currents and the currents in $c=1$ string theory has already been discussed
in the bulk of the paper. It remains to understand the currents $R, R^{\ast}$ and $Y,Y^\ast$. 
The operator $R$ is a chiral primary operator in the untwisted theory with conformal
dimension and R-charge $(\Delta,Q)= (\frac{3}{2},3)$. From equation
\eqref{Vtachyon} in the bulk of the paper, it follows that it can be mapped to the left-moving part of the operator $\CV_{-\frac{3}{2}}$, which has the $c=1$ representation
\be\label{omega}
R(z)\mapsto c\,e^{-i\sqrt{2} X(z)} \,.
\ee
From the definition of the operator
$Y(z)$ in \eqref{fullOPE}, and the mapping of the supercurrent $G^-(z)$ in \eqref{bghosts}, we conclude that
\be\label{y}
Y(z)= G^{-}_{-1}\cdot R(z) \mapsto \b(z)\, e^{-i\sqrt{2} X(z)} \,.
\ee
This is the integrand of the operator denoted $K^-(z)$ we introduced in \eqref{kminus}. It plays a crucial role in constructing representatives of the whole $c=1$ string cohomology.
%Since $\b$ is put to a constant $\sqrt{\mu}$ in the KPZ formulation of $c=1$ string theory, we see that $Y(z)$ coincides with one of the $SU(2)$ current algebra generators
%(for the $c=1$ string at self-dual radius). 
%Moreover, we see that  the current $R$ is composed of both the $SU(2)$ current and a ghost part. 
%We thus see that the twisted $\CN=2$ algebra of the bosonic string can also be extended
% by including the currents in \eqref{omega} and \eqref{y}.
We have thus identified a small subalgebra of the large symmetry group of the $c=1$ string with higher spin currents that are already present in the full string theory.
\end{appendix}

\bibliographystyle{unsrt}

\begin{thebibliography}{10}


%\cite{Mukhi:1993zb}
\bibitem{Mukhi}
  S.~Mukhi and C.~Vafa,
 ``Two-dimensional black hole as a topological coset model of c = 1 string theory,''  Nucl.\ Phys.\ B {\bf 407}, 667 (1993)  [arXiv:hep-th/9301083].
%%CITATION = HEP-TH 9301083;%%

%%\cite{Aganagic:2003qj}
%\bibitem{Aganagic}
%  M.~Aganagic, R.~Dijkgraaf, A.~Klemm, M.~Marino and C.~Vafa,
%  ``Topological strings and integrable hierarchies,'' [arXiv:hep-th/0312085].
%  %%CITATION = HEP-TH 0312085;%%

%\cite{Ahn:2004qb}
\bibitem{Ahn}
  C.~Ahn, M.~Stanishkov and M.~Yamamoto,
  ``ZZ-branes of N = 2 super-Liouville theory,''
  JHEP {\bf 0407}, 057 (2004) [arXiv:hep-th/0405274].
  %%CITATION = HEP-TH 0405274;%%

%\cite{Argurio:2000tb}
\bibitem{Argurio:2000tb}
  R.~Argurio, A.~Giveon and A.~Shomer,
  ``Superstrings on AdS(3) and symmetric products,''
  JHEP {\bf 0012}, 003 (2000)
  [arXiv:hep-th/0009242].
  %%CITATION = HEP-TH 0009242;%%

%\cite{Fotopoulos:2005cn}
\bibitem{Fotopoulos:2005cn}
A.~Fotopoulos, V.~Niarchos and N.~Prezas,
%``D-branes and SQCD in non-critical superstring theory,''
JHEP {\bf 0510}, 081 (2005)
[arXiv:hep-th/0504010].
%%CITATION = HEP-TH 0504010;%%

  %\cite{Ashok:2005py}
\bibitem{Ashok}
  S.~K.~Ashok, S.~Murthy and J.~Troost,
  ``D-branes in non-critical superstrings and minimal super Yang-Mills in
  various dimensions,''
  arXiv:hep-th/0504079.
  %%CITATION = HEP-TH 0504079;%%

%\cite{Ashok2}
\bibitem{Ashok2}
S.~K.~Ashok and J.~Troost,
 ``The topological cigar observables,''
 arXiv:hep-th/0604020.
  %%CITATION = HEP-TH 0604020;%%


%\cite{Bershadsky:1993cx}
\bibitem{BCOV}
  M.~Bershadsky, S.~Cecotti, H.~Ooguri and C.~Vafa,
 ``Kodaira-Spencer theory of gravity and exact results for quantum string amplitudes,''
  Commun.\ Math.\ Phys.\  {\bf 165}, 311 (1994)  [arXiv:hep-th/9309140].
  %%CITATION = HEP-TH 9309140;%%

 %\cite{Dijkgraaf:1991ba}
\bibitem{Dijkgraaf}
  R.~Dijkgraaf, H.~L.~Verlinde and E.~P.~Verlinde,
 ``String propagation in a black hole geometry,''
  Nucl.\ Phys.\ B {\bf 371}, 269 (1992).
  %%CITATION = NUPHA,B371,269;%%

%\cite{Eguchi:2003ik}
\bibitem{Eguchi}
  T.~Eguchi and Y.~Sugawara,
  ``Modular bootstrap for boundary N = 2 Liouville theory,''
  JHEP {\bf 0401}, 025 (2004) [arXiv:hep-th/0311141].
  %%CITATION = HEP-TH 0311141;%%

%\cite{Fateev:2000ik}
\bibitem{Fateev}
  V.~Fateev, A.~B.~Zamolodchikov and A.~B.~Zamolodchikov,
  ``Boundary Liouville field theory. I: Boundary state and boundary  two-point function,'' [arXiv:hep-th/0001012].
 %%CITATION = HEP-TH 0001012;%%

%\cite{Frenkel:1992ex}
\bibitem{Frenkel}
  E.~Frenkel,
  ``Determinant formulas for the free field representations of the Virasoro 
and Kac-Moody algebras,''
  Phys.\ Lett.\ B {\bf 286} (1992) 71.
  %%CITATION = PHLTA,B286,71;%%

 %\cite{GhoshalMM}
\bibitem{GhoshalMM}
  D.~Ghoshal, S.~Mukhi and S.~Murthy,
  ``Liouville D-branes in two-dimensional strings and open string field
  theory,''
  JHEP {\bf 0411}, 027 (2004)
  [arXiv:hep-th/0406106].
  %%CITATION = HEP-TH 0406106;%%

 %\cite{Ghoshal:1995wm}
\bibitem{Ghoshal}
  D.~Ghoshal and C.~Vafa,
  ``C = 1 string as the topological theory of the conifold,''
  Nucl.\ Phys.\ B {\bf 453}, 121 (1995)
  [arXiv:hep-th/9506122].
  %%CITATION = HEP-TH 9506122;%%

%\cite{Giveon:2003wn}
\bibitem{Giveon}
  A.~Giveon, A.~Konechny, A.~Pakman and A.~Sever,
  ``Type 0 strings in a 2-d black hole,''   JHEP {\bf 0310}, 025 (2003) [arXiv:hep-th/0309056].
%%CITATION = HEP-TH 0309056;%%

%%\cite{Gopakumar:1998vy}
%\bibitem{Gopakumar}
%  R.~Gopakumar and C.~Vafa,
%  ``Topological gravity as large N topological gauge theory,'' Adv.\ Theor.\ Math.\ Phys.\  {\bf 2}, 413 (1998)   [arXiv:hep-th/9802016].
%  %%CITATION = HEP-TH 9802016;%%

%\cite{Hanany:1994ya}
\bibitem{Hanany:1994ya}
A.~Hanany and Y.~Oz,
%``c = 1 discrete states correlators via W(1+infinity) constraints,''
Phys.\ Lett.\ B {\bf 347}, 255 (1995)
[arXiv:hep-th/9410157].
%%CITATION = HEP-TH 9410157;%%

%\cite{Hanany:2002ev}
\bibitem{Hanany}
  A.~Hanany, N.~Prezas and J.~Troost,
 ``The partition function of the two-dimensional black hole conformal  field theory,''
  JHEP {\bf 0204}, 014 (2002)
  [arXiv:hep-th/0202129].
  %%CITATION = HEP-TH 0202129;%%

%\cite{Hosomichi:2004ph}
\bibitem{hosomichi}
  K.~Hosomichi,
  ``N = 2 Liouville theory with boundary,''
  arXiv:hep-th/0408172.
  %%CITATION = HEP-TH 0408172;%%

%\cite{Israel:2004jt}
\bibitem{janbranes}
D.~Israel, A.~Pakman and J.~Troost,
``D-branes in N = 2 Liouville theory and its mirror,'' [arXiv:hep-th/0405259].
%\cite{Israel:2004jt}

%\cite{Kac:1979fz}
\bibitem{Kac}
  V.~G.~Kac and D.~A.~Kazhdan,
  ``Structure Of Representations With Highest Weight Of Infinite Dimensional Lie Algebras,''
  Adv.\ Math.\  {\bf 34}, 97 (1979).
  %%CITATION = ADMTA,34,97;%%

%\cite{Kazama:1988qp}
\bibitem{Kazama}
  Y.~Kazama and H.~Suzuki,
  ``New N=2 Superconformal Field Theories And Superstring Compactification,''   Nucl.\ Phys.\ B {\bf 321}, 232 (1989).
%%CITATION = NUPHA,B321,232;%%

%\cite{Klebanov:2003km}
\bibitem{Klebanov:2003km}
  I.~R.~Klebanov, J.~Maldacena and N.~Seiberg,
  ``D-brane decay in two-dimensional string theory,''
  JHEP {\bf 0307}, 045 (2003)
  [arXiv:hep-th/0305159].
  %%CITATION = HEP-TH 0305159;%%

%\cite{Knizhnik:1988ak}
\bibitem{Knizhnik}
  V.~G.~Knizhnik, A.~M.~Polyakov and A.~B.~Zamolodchikov,
 ``Fractal Structure Of 2d-Quantum Gravity,'' Mod.\ Phys.\ Lett.\ A {\bf 3}, 819 (1988).
  %%CITATION = MPLAE,A3,819;%%

%\cite{Maldacena:2000hw}
\bibitem{Maldacena}
  J.~M.~Maldacena and H.~Ooguri, ``Strings in AdS(3) and SL(2,R) WZW model. I,''
  J.\ Math.\ Phys.\  {\bf 42}, 2929 (2001) [arXiv:hep-th/0001053].
  %%CITATION = HEP-TH 0001053;%%

%\cite{Marcus:1992yx}
\bibitem{Marcus}
  N.~Marcus and Y.~Oz,
 ``Discrete states of 2-D string theory in Polyakov's light cone gauge,''
  Nucl.\ Phys.\ B {\bf 392}, 281 (1993)
  [arXiv:hep-th/9206014].
  %%CITATION = HEP-TH 9206014;%%

%\cite{McGreevy:2003kb}
\bibitem{McGreevy}
  J.~McGreevy and H.~L.~Verlinde,
  ``Strings from tachyons: The c = 1 matrix reloaded,''
  JHEP {\bf 0312}, 054 (2003)
  [arXiv:hep-th/0304224].
  %%CITATION = HEP-TH 0304224;%%

%\cite{Murthy:2003es}
\bibitem{Murthy}
  S.~Murthy, ``Notes on non-critical superstrings in various dimensions,''
  JHEP {\bf 0311}, 056 (2003)
  [arXiv:hep-th/0305197].
  %%CITATION = HEP-TH 0305197;%%

%\cite{Nakamura:2005sm}
\bibitem{Nakamura}
  S.~Nakamura and V.~Niarchos,
  ``Notes on the S-matrix of bosonic and topological non-critical strings,''
  arXiv:hep-th/0507252.
  %%CITATION = HEP-TH 0507252;%%

%\cite{Odake:1988bh}
\bibitem{Odake:1988bh}
S.~Odake,
``Extension Of N=2 Superconformal Algebra And Calabi-Yau Compactification,''
Mod.\ Phys.\ Lett.\ A {\bf 4} (1989) 557.
%%CITATION = MPLAE,A4,557;%%

%\cite{Ohta:1993eh}
\bibitem{Ohta1}
  N.~Ohta and H.~Suzuki,
  ``Bosonization of a topological coset model and noncritical string theory,''
  Mod.\ Phys.\ Lett.\ A {\bf 9}, 541 (1994)
  [arXiv:hep-th/9310180].
  %%CITATION = HEP-TH 9310180;%%

%\cite{Itoh:1993mt}
\bibitem{Ohta2}
  K.~Itoh, H.~Kunitomo, N.~Ohta and M.~Sakaguchi,
  ``BRST Analysis of physical states in two-dimensional black hole,''
  Phys.\ Rev.\ D {\bf 48}, 3793 (1993)
  [arXiv:hep-th/9305179].
  %%CITATION = HEP-TH 9305179;%%


%%\cite{Ooguri:1996ck}
%\bibitem{Ooguri}
%  H.~Ooguri, Y.~Oz and Z.~Yin,
%  ``D-branes on Calabi-Yau spaces and their mirrors,''
%  Nucl.\ Phys.\ B {\bf 477}, 407 (1996) [arXiv:hep-th/9606112].
%  %%CITATION = HEP-TH 9606112;%%

%\cite{Pakman:2003cu}
\bibitem{Pakman}
  A.~Pakman,
  ``Unitarity of supersymmetric SL(2,R)/U(1) and no-ghost theorem for fermionic strings in AdS(3) x N,''
  JHEP {\bf 0301}, 077 (2003)
  [arXiv:hep-th/0301110].
  %%CITATION = HEP-TH 0301110;%%

%\cite{Polyakov:1987zb}
\bibitem{Polyakov}
  A.~M.~Polyakov, ``Quantum Gravity In Two-Dimensions,''
  Mod.\ Phys.\ Lett.\ A {\bf 2}, 893 (1987).
  %%CITATION = MPLAE,A2,893;%%

%\cite{Ponsot:2001gt}
\bibitem{Ponsot}
  B.~Ponsot, V.~Schomerus and J.~Teschner,
  ``Branes in the Euclidean AdS(3),'' JHEP {\bf 0202}, 016 (2002) [arXiv:hep-th/0112198].
%%CITATION = HEP-TH 0112198;%%

\bibitem{Polchinski}
J.~Polchinski, String Theory, Vol. 1 Cambridge University Press.

%\cite{Rastelli:2005ph}
\bibitem{Rastelli} 
L.~Rastelli and M.~Wijnholt, ``Minimal AdS(3),'' arXiv:hep-th/0507037.
  %%CITATION = HEP-TH 0507037;%%

%\cite{Recknagel}
\bibitem{Recknagel}
  A.~Recknagel and V.~Schomerus,
  ``Boundary deformation theory and moduli spaces of D-branes,''
  Nucl.\ Phys.\ B {\bf 545}, 233 (1999)
  [arXiv:hep-th/9811237].
  %%CITATION = HEP-TH 9811237;%%

%\cite{Ribault:2003ss}
\bibitem{Ribault}
  S.~Ribault and V.~Schomerus,
  ``Branes in the 2-D black hole,''  JHEP {\bf 0402}, 019 (2004)  [arXiv:hep-th/0310024].
%%CITATION = HEP-TH 0310024;%%

%\cite{Ribault:2005wp}
\bibitem{RibaultTeschner}
  S.~Ribault and J.~Teschner,
  ``H(3)+ correlators from Liouville theory,''
  JHEP {\bf 0506}, 014 (2005)
  [arXiv:hep-th/0502048].
  %%CITATION = HEP-TH 0502048;%%

%%\cite{Schomerus:2002dc}
%\bibitem{Schomerus:2002dc}
%  V.~Schomerus,
%  ``Lectures on branes in curved backgrounds,''
%  Class.\ Quant.\ Grav.\  {\bf 19}, 5781 (2002)
%  [arXiv:hep-th/0209241].
%  %%CITATION = HEP-TH 0209241;%%

%%\cite{Seiberg:2003nm}
%\bibitem{seibergshih}
%  N.~Seiberg and D.~Shih,
% ``Branes, rings and matrix models in minimal (super)string theory,''
%  JHEP {\bf 0402}, 021 (2004) [arXiv:hep-th/0312170].
%  %%CITATION = HEP-TH 0312170;%%

%\cite{Stoyanovsky:2000pg}
\bibitem{Stoyanovsky}
  A.~V.~Stoyanovsky,
  ``A relation between the Knizhnik--Zamolodchikov and Belavin--Polyakov--Zamolodchikov systems of partial differential equations,''
  arXiv:math-ph/0012013.
  %%CITATION = MATH-PH 0012013;%%

%\cite{Takayanagi:2005yb}
\bibitem{Takayanagi}
  T.~Takayanagi,
  ``$c < 1$ string from two dimensional black holes,''
  arXiv:hep-th/0503237.
%%CITATION = HEP-TH 0503237;%%

%\cite{Teschner:1997ft}
\bibitem{Teschnerpapers}
  J.~Teschner,  ``On structure constants and fusion rules in the SL(2,C)/SU(2) WZNW  model,''
  Nucl.\ Phys.\ B {\bf 546}, 390 (1999), [arXiv : hep-th/9712256].\ , 
  ``The mini-superspace limit of the SL(2,C)/SU(2) WZNW model,''
  Nucl.\ Phys.\ B {\bf 546}, 369 (1999)
  [arXiv : hep-th/9712258]. \ , 
  ``Operator product expansion and factorization in the H-3+ WZNW model,''
  Nucl.\ Phys.\ B {\bf 571}, 555 (2000)
  [arXiv : hep-th/9906215].
  %%CITATION = HEP-TH 9712256;%%

%\cite{Teschner:2000md}
\bibitem{Teschner}
  J.~Teschner,
  ``Remarks on Liouville theory with boundary,''
  arXiv:hep-th/0009138.
  %%CITATION = HEP-TH 0009138;%%

%\cite{Witten:1991zd}
\bibitem{Witten}
  E.~Witten,
 ``Ground ring of two-dimensional string theory,''
  Nucl.\ Phys.\ B {\bf 373}, 187 (1992)
  [arXiv:hep-th/9108004].
  %%CITATION = HEP-TH 9108004;%%

%\cite{Witten:1992yj}
\bibitem{WittenZ}
 E.~Witten and B.~Zwiebach,
 ``Algebraic structures and differential geometry in $2-D$ string theory,''
  Nucl.\ Phys.\ B {\bf 377}, 55 (1992) [arXiv:hep-th/9201056].
  %%CITATION = HEP-TH 9201056;%%

%\cite{Zamolodchikov:2001ah}
\bibitem{Zamolodchikov}
  A.~B.~Zamolodchikov and A.~B.~Zamolodchikov,
  ``Liouville field theory on a pseudosphere,''  [arXiv:hep-th/0101152].
%%CITATION = HEP-TH 0101152;%%

%%\cite{Zamolodchikov:2005jb}
%\bibitem{AlbZ}
%  A.~Zamolodchikov,
% ``Perturbed conformal field theory on fluctuating sphere,''
%  arXiv:hep-th/0508044.
%  %%CITATION = HEP-TH 0508044;%%



\end{thebibliography}

\end{document}